\documentclass[useAMS,usenatbib]{mn2e}

\usepackage{epsfig}
\usepackage{amsmath, amssymb,bm}

\title[Supernova Kicks and Misaligned Be Star Binaries]{Supernova
  Kicks and Misaligned Be Star Binaries}

\author[R. G. Martin, C. A. Tout and J. E. Pringle]{Rebecca G. Martin,
  Christopher A. Tout and J. E. Pringle\\ University of Cambridge,
  Institute of Astronomy, The Observatories, Madingley Road, Cambridge
  CB3 0HA\\}
\begin{document}

\date{}

\pagerange{\pageref{firstpage}--\pageref{lastpage}} 
\pubyear{2007}
\maketitle

\label{firstpage}

\begin{abstract}

  Be~stars are rapidly spinning B~stars surrounded by an outflowing
  disc of gas in Keplerian rotation.  Be~star/X-ray binary systems
  contain a Be~star and a neutron star. They are found to have
  non-zero eccentricities and there is evidence that some systems have
  a misalignment between the spin axis of the star and the spin axis
  of the binary orbit. The eccentricities in these systems are be
  caused by a kick to the neutron star during the supernova that
  formed it. Such kicks would also give rise to misalignments. In this
  paper we investigate the extent to which the same kick distribution
  can give rise to both the observed eccentricity distribution and the
  observed misalignments. We find that a Maxwellian distribution of
  velocity kicks with a low velocity dispersion, $\sigma_k \approx
  15\rm \, km\, s^{-1}$, is consistent with the observed eccentricity
  distribution but is hard to reconcile with the observed
  misalignments, typically $i \ge 25^\circ$.  Alternatively a higher
  velocity kick distribution, $\sigma_k = 265 \rm \, km\, s^{-1}$, is
  consistent with the observed misalignments but not with the observed
  eccentricities, unless post-supernova circularisation of the binary
  orbits has taken place.  We discuss briefly how this might be
  achieved.

\end{abstract}


\section{Introduction}

Be~stars were discovered by \cite{Secchi} who observed emission lines
in $\gamma$~Cas.  These stars are rapidly rotating at about 70 per
cent of their break up velocity \citep{P96}. In fact, relative to
their break up velocities, they are the fastest rotating bodies
observed.  They are early type main-sequence stars which have shown
H$\alpha$ in emission at least once. They are variable in brightness
and spectra which show broad HeI absorption and emission at either
visual or UV wavelengths.  It is found that the emission, and so
presumably the discs, are only temporary and so Be stars become B
stars and vice versa.

If the disc is viewed edge on the Be~star is seen as a shell star.
The spectra then show Balmer emission with sharp absorption cores,
narrow absorption lines of ionized metals and broad HeI absorption.
Be-star discs vanish and re-appear on timescales of a few hundred
days.  \cite{D88} studied Balmer emission line profiles and concluded
that the envelopes surrounding Be~stars are in Keplerian motion within
the disc. The disc shows optical and IR emission lines and an IR
continuum excess.

The stars $\gamma$~Cas and 59~Cyg have shown two successive shell
events.  These were associated with a remarkably synchronous
quasi-cyclic variation of the emission line width in all observed
emission lines that has been called spectacular variation \citep{H98}.
The change in emission line width removes the correlation between the
projected surface velocity, $v\sin i$, and the FWHM (full width half
maximum) and so a circumstellar equatorial disc fails to explain the
spectacular variations.  The emission lines and shell Be~stars are
explained by differences in disc inclination to the line of sight, so
transitions between the two were not expected.  \cite{H98} explains
the spectacular variations by a Keplerian disc which is somehow tilted
with respect to the equatorial plane of the star.  The variation in
emission line widths and profile shapes are then due to the precession
of the disc.  He suggests that the sequence of alternating
shell-phases and narrow single peak phases is due to the variation in
disc inclination caused by precession.  The idea that a disc might
change its inclination to the line of sight is borne out by
observations of 28 Tau (Pleione) by \cite{hirata07}. This star also
changes between B star, Be star and shell star and in this star the
intrinsic polarization angle changes in phase with these variations.
The cause of the precession is also not clear but \cite{H98} suggested
that it might be induced by tides from a binary companion.  The two
systems $\gamma$~Cas and 59~Cyg are binary. The system 28~Tau appears
to show radial velocity variations, although there is no confirmed
orbital period \citep{riv06}. In both $\gamma$~Cas and 28~Tau the
misalignment angle between the stellar equator and the disc/orbital
plane is thought to be around $25^\circ$ \citep{H98, hirata07}.

In further support of this possibility, we note that the B-star binary
PSR J0045--7319 has a spin-orbit misalignment suggested by its orbital
plane precession \citep{K96,L95}.  This misalignment in a B-star orbit
means that misalignment in Be stars in not uncommon. In this case the
B star rotates retrogradely with respect to the orbit \citep{L96a}.

The standard model for Be-star discs is that they are decretion discs
with the mass expelled from the neighbourhood of the Be~star itself
\citep{cass02}. In this case we expect the plane of the inner disc to
be aligned with the spin axis of the Be~star. There are two reasons
for the disc to be found at an angle discussed in the literature.
First \cite{porter98} suggested that the disc warping and precession
might be caused by a radiation-induced instability \citep{Pringle96}.
Secondly, as we reported above, it is widely suggested that the warp
and precession are caused by a misalignment between the spin axis of
the Be~star and the orbit of the binary companion. If the disc is a
decretion disc we expect the inner edge to be aligned with the
equatorial plane of the B~star and the outer edge to tend to be
tidally aligned with the orbital plane. Thus there must be a warp at
some radius in the disc.  Here we focus on this second possibility and
concentrate on the Be/X-ray binaries in which the companion stars are
neutron stars.

Neutron stars, observed as radio pulsars, have space velocities much
greater than their progenitors \citep{GO70}. The accepted explanation
for this is that supernova explosions are asymmetric and give very
large kicks to the newly formed neutron stars \citep{S70,S78}. Indeed
some supernova remnants show evidence for asymmetric explosions
\citep{MWK95, A95}. Thus any system that contains a neutron star could
have had a supernova kick.  It is also found that Be/X-ray binaries
have high eccentricities that cannot be explained without supernova
kicks \citep{VH95}.

In a Be-star binary system, prior to the supernova in which the core
of its companion collapses to a neutron star, we expect the
Be-star spin to aligned with the orbit and for the orbit to be
circular. The kick from the supernova has two effects, it makes the
orbit eccentric, and perhaps even unbinds it, and it misaligns the
orbit with the spin axis of the Be~star. Thus information about the
distribution of eccentricities in Be-star systems can in principle
give us information about the distribution of spin--orbit
misalignments.

\cite{LL94} analysed the known pulsar velocities and concluded that
they were born with a mean speed of about $450\,\rm km\,s^{-1}$.
\cite{HP97} considered the selection effects, as a result of the flux
limits, of the pulsar surveys and the accuracy of the proper motion
determinations and found a mean birth speed of around $250-300\,\rm
km\,s^{-1}$. This is consistent with a Maxwellian distribution with
$\sigma_{\rm k}=190\,\rm km\,s^{-1}$ which has a mean velocity of
$\langle v_k \rangle = 303\,\rm km\,s^{-1}$.  More recently
\cite{arzou} found a best fitting distribution with two Maxwellian
components, one for $40$ per cent of the pulsars with $\sigma_{\rm
  k1}=90\,\rm km\,s^{-1}$ and the other with $\sigma_{\rm k2}=500\,\rm
km\,s^{-1}$.

Subsequently \cite{hobbs05} extended the work of \cite{LL94} with a
much larger sample of single radio pulsars and claimed that the kick
distribution is consistent with a single Maxwellian with $\sigma_{\rm
  k}=265\,\rm km\,s^{-1}$ without a significant low-velocity
component. That such a kick distribution cannot reproduce the
period--eccentricity distribution of Be/X-ray binaries has already
been noticed and discussed by \cite{pfahl02} and \cite{vanderh07}.
They concluded that a bimodal kick distribution is needed with a
low-velocity component ($\sigma_{\rm k}<50\,\rm km\,s^{-1}$) when the
supernova occurs in a binary system.  The idea that there might be two
types of collapse and therefore two types of supernova kick was
originally proposed by \cite{katz}. \cite{pod04} also find evidence
for a two-component kick distribution with low-velocity kicks mainly
in close binaries.  \cite{kramer08} give an extensive discussion of
the second kick in the double pulsar J0737--3039 and conclude that it
probably had to be small.  Theoretical studies such as those by
\cite{scheck06} and \cite{kita06} do not yet throw much light on this
but it may be that there are more lower mass supernovae, with smaller
kicks, in binary systems or that the binary orbit quenches the
hydrodynamic instabilities which lead to a very asymmetric explosion.

In a binary system, if the kick is too strong, the system does not
remain bound.  Even a small velocity kick can lead to a large
eccentricity and inclination between the old and new orbits
\citep{BP95}.  We do not know how the angular momentum of the remnant
is also affected by the supernova kick so we do not know how much of
the misalignment of the neutron star now is caused by the orbital
inclination.  We can however expect the companion to continue spinning
aligned with the pre-supernova orbit immediately after the explosion.
There are several binary systems with neutron star companions that are
observed to be misaligned.

\cite{BP95} investigated some of the effects of high supernova kick
velocities on the orbital parameters of post-supernova neutron-star
binaries.  Here we look at a variety of velocity kick distributions
and consider the implications for the distribution of the inclinations
between the orbit before and after the kick.  After ensuring that we
can reproduce the work of \cite{BP95} we model Be-star systems with
our preferred distributions for their progenitors including a range of
masses.  There is somewhat more data now available for comparison in
the period--eccentricity plane and we find that, though all systems
can be formed with our models, they tend to be more circular than
expected with kicks distributed according to \citet{hobbs05}. We
investigate what kick distribution could lead to the observed
eccentricity distribution and also how the eccentricity distribution
might have changed since the supernova. We consider three types of
kick distribution, a single peaked Maxwellian velocity kick
distribution, kicks which are direction limited and a double
Maxwellian distribution.



\section{Misalignment Probability Distribution}

\begin{figure}
\begin{center}
  \epsfxsize=8.4cm \epsfbox{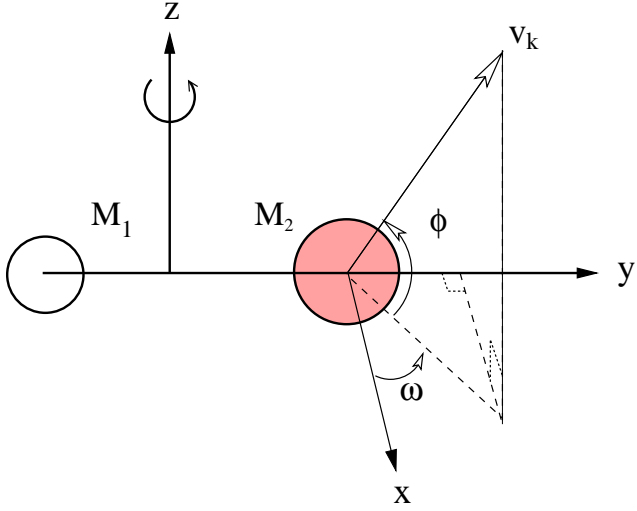}
  \caption[] {The system before the supernova. The two
    stars of mass $M_1$ and $M_2$ are in a circular orbit about their
    centre of mass at the origin. The orbital angular momentum is in
    the $z$-direction and the orbit is in the $xy$-plane.  When star 2
    explodes as a supernova it is travelling in the direction of the
    {\it negative} $x$-axis with speed $v_{\rm orb}$ relative to star
    1.  It receives a kick of velocity $v_{\rm k}$ at an angle $\phi$
    to the plane of the binary orbit. The angle between the projection
    of the velocity kick on to the binary plane and the $x$-direction
    is $\omega$.}
\label{geometry}
\end{center}
\end{figure}

In this section we consider the effect of a velocity kick on the
orbital inclination.  We start with a binary in a circular orbit.  One
star then has an asymmetric supernova explosion which gives it a kick
with velocity $0\le v_{\rm k}<\infty$ in a direction given by the
angle $\phi$ out of the binary plane ($-\pi/2\le\phi\le\pi/2$) and an
angle between the direction opposite to the instantaneous velocity of
the star and the projection of the velocity kick into the binary
orbital plane of $0\le\omega<2\pi$ (see Fig.~\ref{geometry}). For now
we assume that no mass is lost.

\begin{figure}
  \epsfxsize=8.4cm \epsfbox{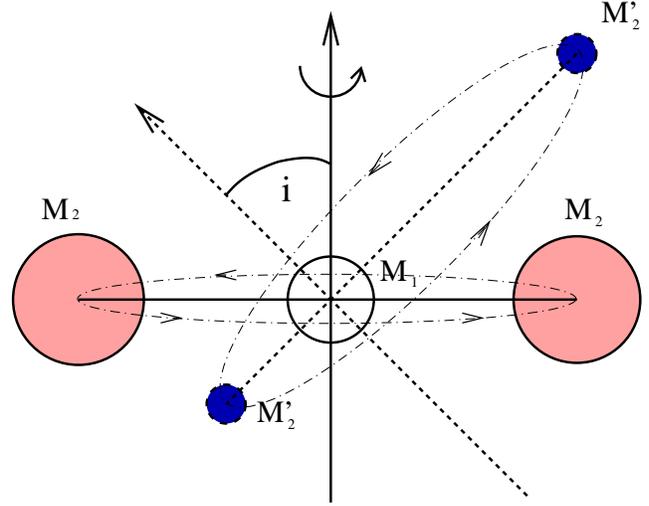}
  \caption[] {The binary system before and after the supernova
    in the frame of the star 1 of mass $M_1$. The mass of star~2 is
    reduced from $M_2$ to $M_2'$ in the supernova and it experiences
    the kick. The straight solid arrow is the orbital angular momentum
    of the system before the supernova. The other solid straight line
    is the pre-supernova diameter of the circular orbit indicated by
    the dot-dashed ellipse.  The dashed arrow is the orbital angular
    momentum after the supernova. The angle $i$ is the angle of
    misalignment between the pre- and post-supernova orbits.  If star
    1 has its spin aligned with the pre-supernova orbit, then the
    angle $i$ also measures the misalignment between the
    post-supernova eccentric orbit and the spin of star 1. The other
    dashed line is the major axis of the post-supernova orbit
    indicated by a dot-dash line.}
\label{geometry2}
\end{figure}

\begin{figure*}
  \centerline{\hbox{ \epsfxsize=8cm  \epsffile{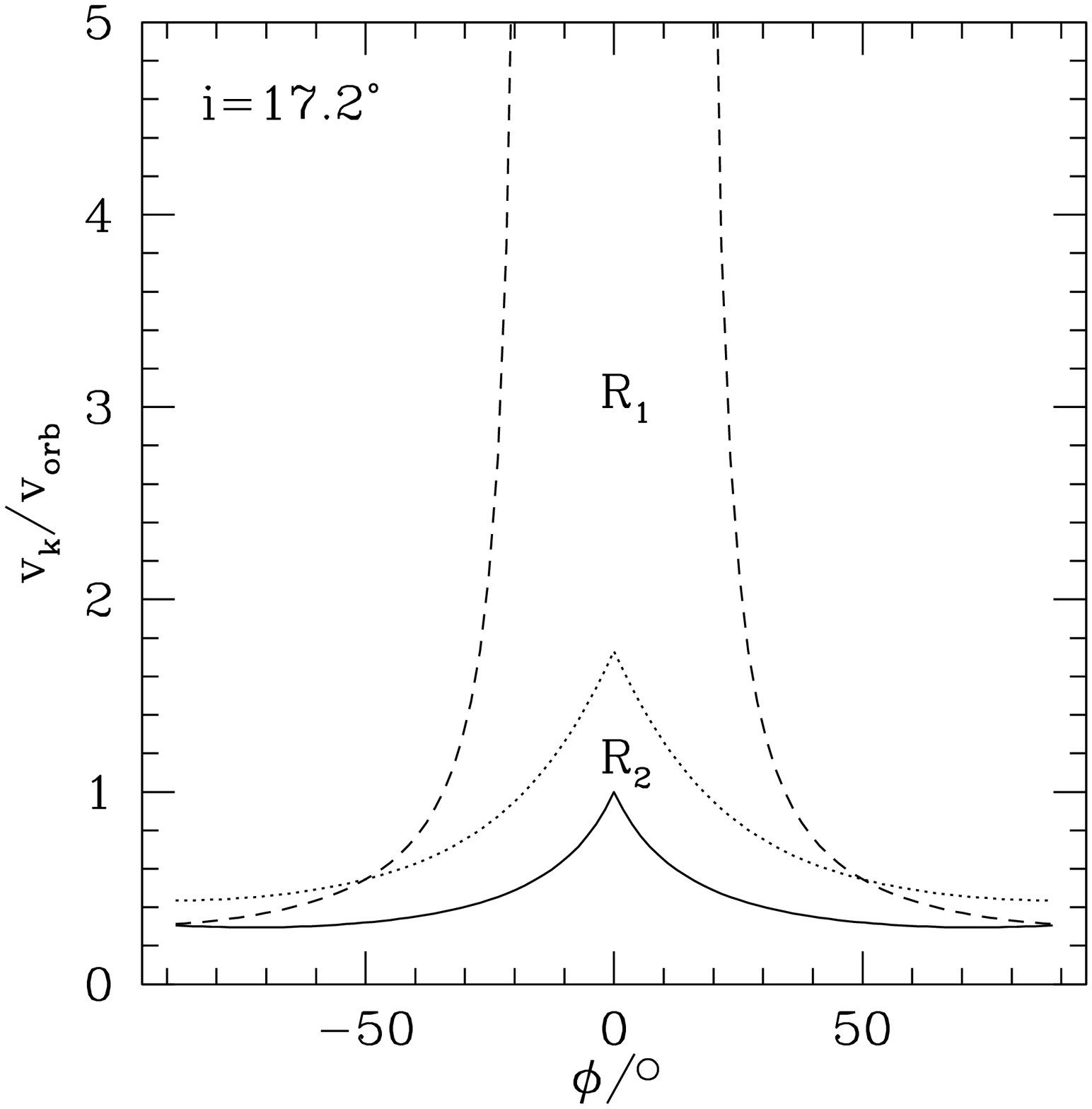}
      \epsfxsize=8cm\epsffile{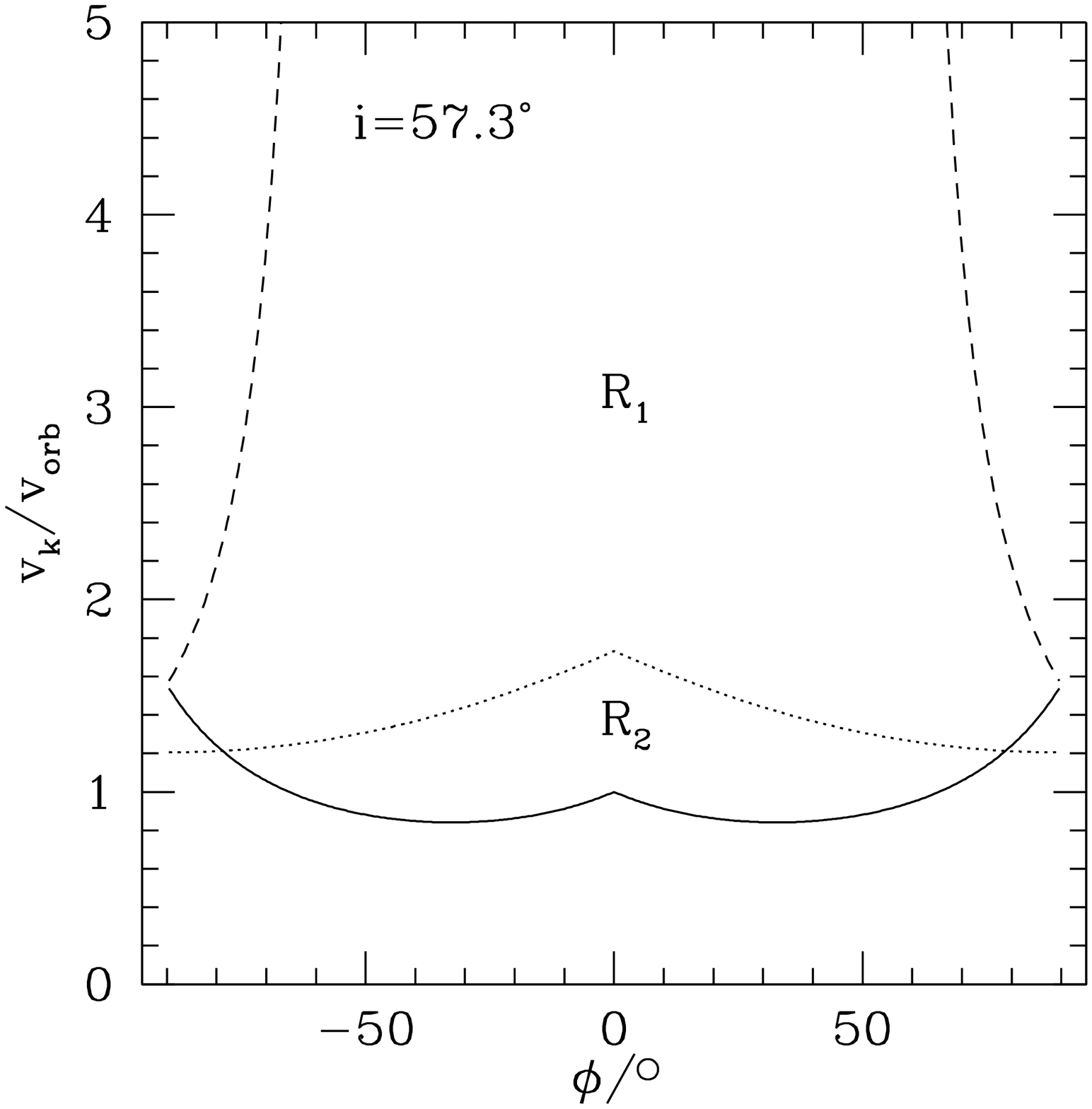} } }
  \centerline{\hbox{ \epsfxsize=8cm \epsffile{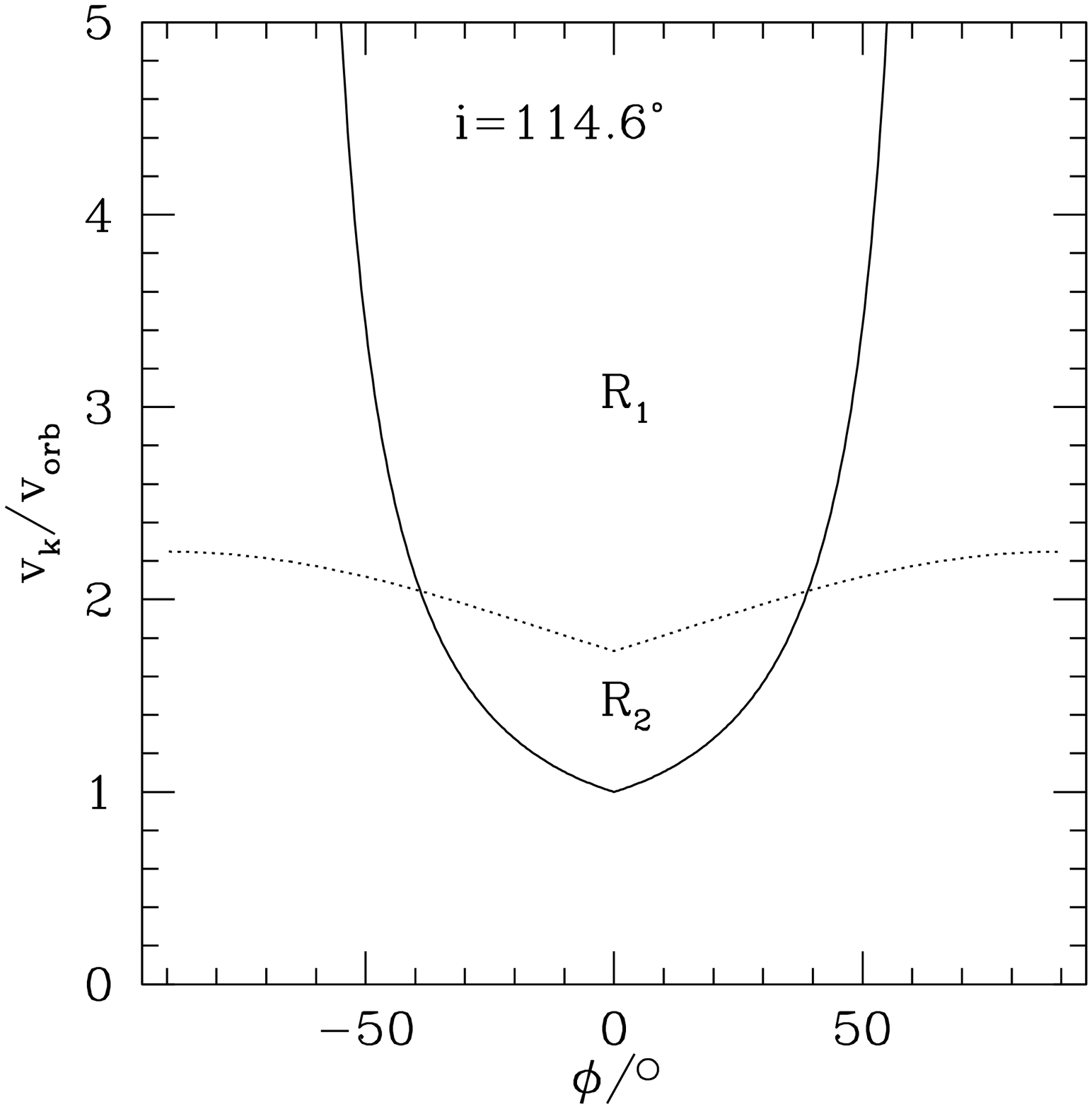}
      \epsfxsize=8cm  \epsffile{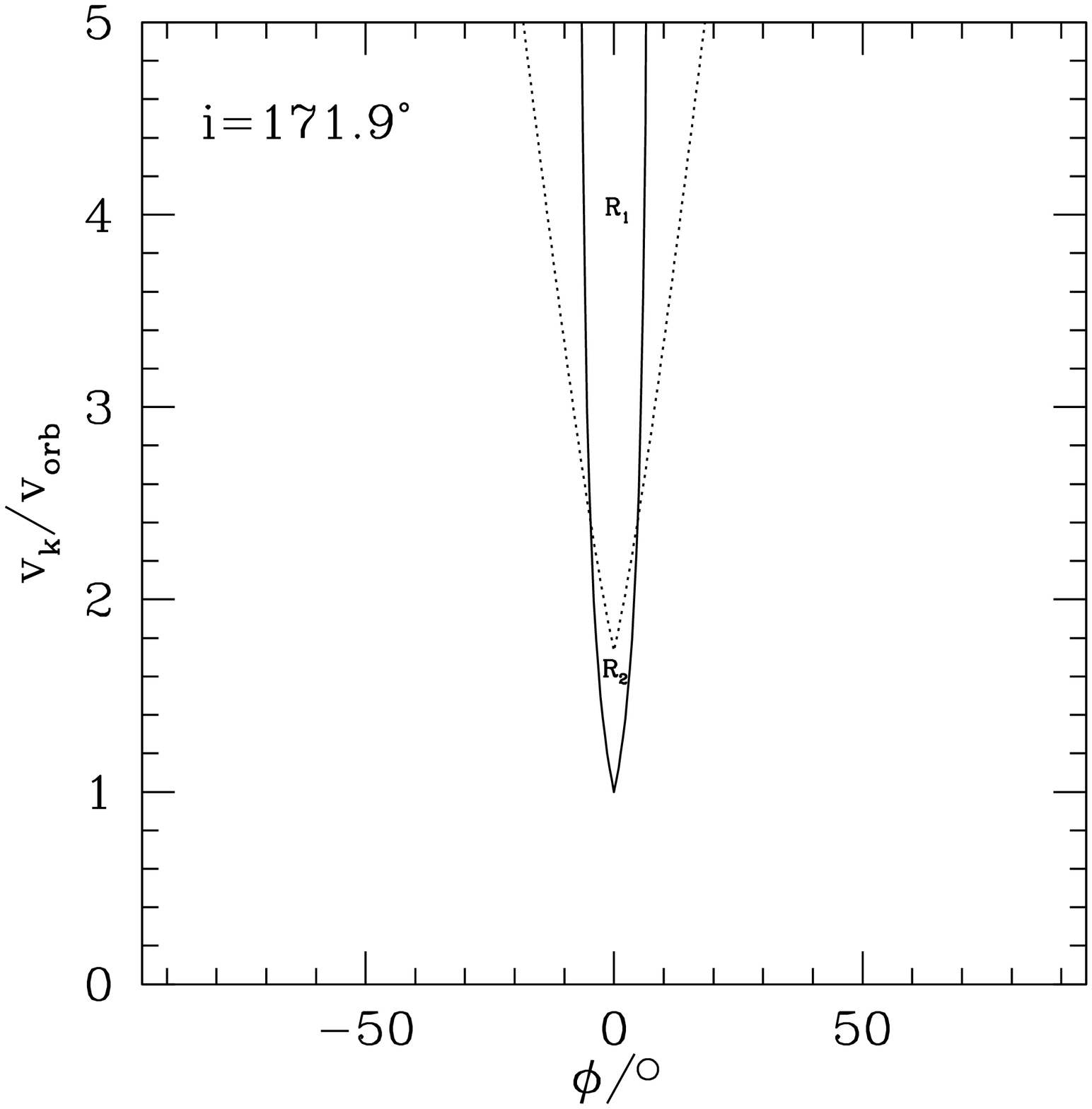}}}
  \caption[] {  Possible combinations of the ratio of star 2's supernova kick
    velocity, $v_{\rm k}$, to the relative orbital velocity, $v_{\rm
      orb}$ and angle $\phi$ between the kick direction and the
    orbital plane for four different inclinations $i$ between the pre-
    and post-supernova orbital planes, top left $i = 17.2^\circ$, top
    right $i = 57.3^\circ$, bottom left $i= 114.6^\circ$ and bottom
    right $i = 171.9^\circ$.  For angles $i > 90^\circ$ the
    post-supernova orbit counter rotates with respect to the spin of
    star~1.  For each value of $\phi$ there is a range of values of
    $v_{\rm k}$ which can give rise to the required misalignment $i$.
    This depends on the angle $\omega$ ($0 \le \omega < 2\pi$) shown
    in Fig.~\ref{geometry}.  In each panel, the solid line corresponds
    to $\omega = 0$ (and thus to the kick velocity $v_+$,
    equation~\ref{v1}) and the dashed line to $\omega = \pi$ (kick
    velocity $v_-$, equation~\ref{v2}).  Note that $\omega = 0$, $\phi
    = 0$ corresponds to a kick directly opposed to the motion of star
    2 and $\omega = \pi$, $\phi=0$ corresponds to a kick in the
    direction of motion of star 2. In the bottom two plots, which
    correspond to post-supernova retrograde motion, it is not possible
    to achieve this with a prograde kick ($\omega =\pi$) for any value
    of $\phi$ so that the dashed lines corresponding to $\omega = \pi$
    are absent. The dotted lines show the maximum velocity kick,
    $v_{\rm bound}$, (equation~\ref{vb}) as a function of $\phi$, for
    which the system remains bound.  Below this line the system
    remains bound after the supernova and above it the binary is
    disrupted. In order to find the probability distribution $P(i)$ of
    the misalignment angle $i$ in equation~(\ref{int}) we integrate in
    ($v_{\rm k}/v_{\rm orb}, \phi$)-space over the regions $R = R_1 +
    R_2$ between the contours of $v_+$ (solid line), $v_-$ (dashed
    line). To find only the probability distribution $P(i)$ for the
    bound systems alone we integrate over the regions $R_2$ only.}
\label{contours}
\end{figure*}

We are interested in the misalignment angle of the system, $i$, after
the supernova kick. This is the angle between the old and new angular
momenta of the orbits (Fig.~\ref{geometry2}). If $0\le i<\pi/2$ then
the system is closer to alignment than counter-alignment and if
$\pi/2<i\le\pi$ it is closer to counter-alignment.  \cite{BP95} find
this angle to be given by
\begin{equation}
\cos i = \frac{v_{\rm orb}-v_{\rm k} \cos\omega \cos \phi}{[v_{\rm k}^2 \sin^2 \phi 
+(v_{\rm orb}-v_{\rm k} \cos\omega \cos \phi)^2]^{\frac{1}{2}}},
\label{hurley}
\end{equation}
where $v_{\rm orb}$ is the initial orbital velocity of the system.  The
relative velocity of the stars after the supernova is
\begin{equation}
v_{\rm n}^2=v_{\rm k}^2+v_{\rm orb}^2-2v_{\rm orb}v_{\rm k}\cos \omega \cos \phi.
\label{vn}
\end{equation}
We can rearrange equation~(\ref{hurley}) to find
\begin{equation}
\cos \omega =\frac{v_{\rm orb}}{v_{\rm k}}\frac{1}{\cos \phi} \pm \frac{\tan \phi}{\tan i}
\label{re}
\end{equation}
when $v_{\rm k} \ne 0$, $\tan i \ne 0$ (so that $i \ne 0,\pi$) and
$\cos \phi \ne 0$ (so that $\phi \ne -\pi/2, \pi/2)$.  Then if $0\le
i<\pi/2$ from equation~(\ref{hurley}) we have $v_{\rm orb}>v_{\rm
  k}\cos \omega \cos \phi$. With equation~(\ref{re}) this corresponds
to
\begin{equation}
\mp \frac{\tan \phi}{\tan i}v_{\rm k}\cos \phi >0
\end{equation}
and because $\tan i>0$ we see 
\begin{equation}
\mp \sin \phi>0.
\end{equation}
Similarly when $\pi/2<i\le\pi$, so that $\tan i<0$, we find the same
condition as above on $\sin \phi$. Now we can rewrite
equation~(\ref{re}) as
\begin{equation}
\cos \omega =\frac{v_{\rm orb}}{v_{\rm k}}\frac{1}{\cos \phi} - \frac{|\tan \phi|}{\tan i}.
\label{re2}
\end{equation}
We consider where this equation has real valued solutions in the
$\phi-v_{\rm k}$ plane. In Fig.~\ref{contours} we plot the locus of
$\cos \omega= 1$ ($\omega=0$),
\begin{equation}
v_+=\frac{v_{\rm orb}}{\cos \phi}\left(1+\frac{|\tan \phi|}{\tan i}\right)^{-1},
\label{v1}
\end{equation}
and the locus of $\cos \omega=-1$ ($\omega=\pi$),
\begin{equation}
v_-=\frac{v_{\rm orb}}{\cos \phi}\left(-1+\frac{|\tan \phi|}{\tan i}\right)^{-1},
\label{v2}
\end{equation}
for four values of $i$.  The region between the $v_+$ and $v_-$
contours is the region where we have real values of $\cos \omega$ and
it represents the combinations of kick parameters which can lead to a
misalignment of the chosen $i$.

\section{Isotropic Maxwellian kick distribution}

As an illustration we apply these results to a simple isotropic
Maxwellian kick distribution. For an isotropic kick distribution the
direction of the kick velocity is uniformly distributed over a sphere
so that the angles defined in Fig.~\ref{geometry} are distributed as
\begin{equation}
P(\omega)d\omega=\frac{1}{2\pi}\, d\omega
\end{equation}
and 
\begin{equation}
P(\phi)d\phi=\cos \phi\, d\phi.
\end{equation}
We here choose the kick speed to have a Maxwellian distribution so
that
\begin{equation}
P(v_{\rm k})dv_{\rm k}=\sqrt{\frac{2}{\pi}}\frac{v_{\rm k}^2}{\sigma_{\rm k}^3}e^{-\frac{v_{\rm k}^2}{2\sigma_{\rm k}^2}}\,dv_{\rm k}
\label{vk},
\end{equation}
where $\sigma_{\rm k}$ is the dispersion of the velocity, and recall
that \cite{hobbs05} find $\sigma_{\rm k}=265\,\rm km\, s^{-1}$. In
sections~\ref{axis} and~\ref{bi} we shall consider alternative
velocity kick distributions.  Because $i=i(v_{\rm k},\phi,\omega)$ its
probability distribution is
\begin{align}
P(i)di=\int_{v_{\rm k}=0}^{\infty} &  \int_{\phi=-\frac{\pi}{2}}^{\frac{\pi}{2}} \int_{\omega=0}^{2\pi}  P(\phi)P(\omega)P(v_{\rm k})\,d\omega\, d\phi \,dv_{\rm k}.
\end{align}
We change variables from $(v_{\rm k},\phi,\omega)$ to $(v_{\rm k},\phi,i)$ and find
\begin{equation}
P(i)di=\int_{i}^{i+di} \!\!\int\!\!\! \int_{R} P(\phi)P(\omega)P(v_{\rm k}) |J|\, d\phi \,dv_{\rm k}\, di,
\end{equation}
where $R=R_1+R_2$ is the region in the $(\phi,v_{\rm k})$ plane where
$\cos \omega$ is real valued. This is illustrated in
Fig.~\ref{contours} for different values of $i$.  Outside of the
region bounded by these curves, the given velocity kick and angle
$\phi$ cannot produce a system misaligned by $i$ because then $|\cos
\omega|>1$. We consider this region for bound systems in
Section~\ref{poss}. The Jacobian, $J$, for the change of variables is
given by
\begin{equation}
dv_{\rm k}\,d\phi\, d\omega = |J| \,dv_{\rm k}\,d\phi\,di,
\end{equation}
where
\begin{equation}
J = \left| \begin{array}{ccc}
\left(\frac{\partial v_{\rm k}}{\partial v_{\rm k}}\right)_{\phi,i} & \left(\frac{\partial v_{\rm k}}{\partial \phi}\right)_{v_{\rm k},i} & \left(\frac{\partial v_{\rm k}}{\partial i}\right)_{v_{\rm k},\phi} \\
\left(\frac{\partial \phi}{\partial v_{\rm k}}\right)_{\phi,i} & \left(\frac{\partial \phi}{\partial \phi}\right)_{v_{\rm k},i} & \left(\frac{\partial \phi}{\partial i}\right)_{v_{\rm k},\phi} \\
 \left(\frac{\partial \omega}{\partial v_{\rm k}}\right)_{\phi,i} & \left(\frac{\partial \omega}{\partial \phi}\right)_{v_{\rm k},i} & \left(\frac{\partial \omega}{\partial i}\right)_{v_{\rm k},\phi} \end{array} \right|.
\end{equation}
Because $\omega$, $\phi$ and $v_{\rm k}$ are independently distributed we find
\begin{equation}
J = \left| \begin{array}{ccc}
1&0&0 \\
0&1&0\\
0&0 & \left(\frac{\partial \omega}{\partial i}\right)_{v_{\rm k},\phi} \end{array} \right|=\left(\frac{\partial \omega}{\partial i}\right)_{v_{\rm k},\phi}.
\end{equation}
We differentiate equation~(\ref{re2}) to find
\begin{equation}
J=\left(\frac{\partial \omega}{\partial i}\right)_{v_{\rm k},\phi}= -\frac{|\tan \phi|}{\sin \omega \sin^2 i}.
\end{equation}
Now we can write down the full probability density distribution for
the misalignment as an integral
\begin{equation}
P(i)di=\sqrt{\frac{2}{\pi^3}}\frac{1}{2 \sigma_k^3}\int_{i}^{i+di} \int \!\!\! \int_{R} I \,d\phi \,dv_{\rm k}\, di,
\label{int}
\end{equation}
where
\begin{equation}
I=v_{\rm k}^2 e^{-\frac{v_{\rm k}^2}{2\sigma_k^2}} \frac{|\sin \phi|}{|\sin \omega| \sin^2 i}
\label{int2}
\end{equation}
and $\omega=\omega(i, v_{\rm k})$ is defined by
equation~(\ref{re2}). 

\begin{figure}
  \epsfxsize=8.4cm \epsfbox{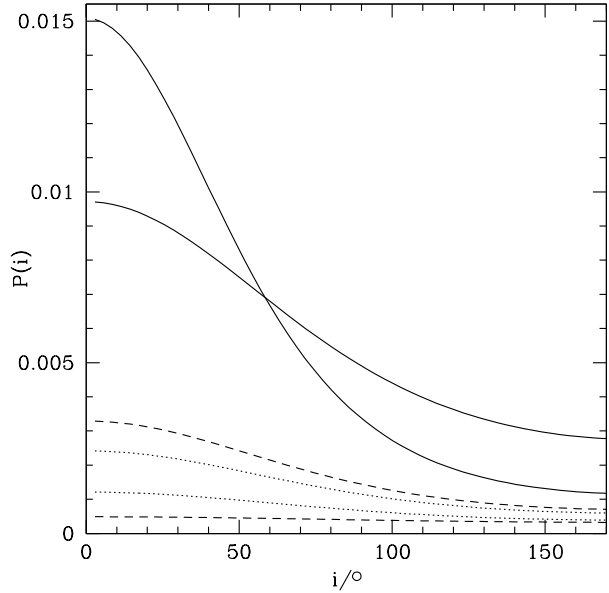}
\caption[] {The probability distribution $P(i)$ of the post-supernova
  misalignment angle $i$ is plotted as a function of $i$ for the case
  when the probability distribution of the kick velocity $v_{\rm k}$
  is a Maxwellian with dispersion $\sigma_{\rm k}$ and no mass
  is lost from the system ($f=0$).  The solid lines correspond to
  integrating over the full area $R = R_1 + R_2$ in
  Fig.~\ref{contours} and so to all systems whether or not they remain
  bound. The curve with the higher value of $P(i=0)$ corresponds to
  $\sigma_{\rm k}/v_{\rm orb} = 1$, and the other one to $\sigma_{\rm
    k}/v_{\rm orb} = 2$. These are normalised so that $\int_0^{180^\circ} P(i)
  \, di = 1$. Note that the distribution with the larger velocity
  spread is more able to produce counter-rotating systems.  The dashed
  lines are the corresponding probability distributions ($\sigma_{\rm
    k}/v_{\rm orb} = 1$, upper; $\sigma_{\rm k}/v_{\rm orb} = 2$,
  lower) with the integral now taken over only the region $R_2$ in
  Fig.~\ref{contours} and so only including those systems which
  remain bound. The change in $P(i)$ for given values of $i$ and
  $\sigma_{\rm k}$ is due to the fraction of systems which become
  unbound.  The dotted lines are the probability distributions $P(i)$
  for systems which remain bound, with $\sigma_{\rm k}/v_{\rm orb} =
  1$, but for which mass loss also occurs. The fractional mass loss is
  $f=0.2$ for the upper dotted line and $f=0.5$ for the lower.}

\label{bound}
\end{figure}

This can now be integrated numerically for specific values of $v_{\rm
  orb}/\sigma_{\rm k}$ to find the probability distribution of the
misalignment angle, $i$, of the system.  We plot this as the solid
lines in Fig.~\ref{bound} for $v_{\rm orb}=\sigma_{\rm k}$ (upper line
on $y$-axis) and $v_{\rm orb}=0.5\,\sigma_{\rm k}$. The higher the
misalignment angle, $i$, the lower its probability of forming. The
higher $v_{\rm orb}/\sigma_{\rm k}$, the lower the relative
probability of a counter-rotating system and the higher the
probability of a system being close to alignment.

\subsection{Bound Systems}
\label{poss}

We have calculated the probability distribution of the misalignment
angle of the system but have not yet considered whether the resulting
system remains bound.  If the kinetic energy of the new system exceeds
the gravitational potential energy then the system does not remain
bound. We no longer have a binary system and the two stars fly apart.

Initially the gravitational energy of the system is
\begin{equation}
E_{\rm grav}=-\frac{GM_1 M_2}{a},
\end{equation}
and the kinetic energy is
\begin{equation}
E_{\rm kin}=\frac{1}{2}\left(\frac{M_1M_2}{M}\right)v_{\rm orb}^2,
\end{equation}
in the centre of mass frame, where $M=M_1+M_2$.  In a circular orbit we
have
\begin{equation}
-E_{\rm kin}=\frac{1}{2}E_{\rm grav}
\end{equation}
and the total energy
\begin{equation}
E_{\rm tot}=-\frac{1}{2}\frac{GM_1 M_2}{a}.
\end{equation}
So
\begin{equation}
E_{\rm grav}=-v_{\rm orb}^2 \frac{M_1 M_2}{M}.
\end{equation}
After the kick the gravitational energy remains the same because in
this case no mass is lost. The kinetic energy becomes
\begin{equation}
E'_{\rm kin}=\frac{1}{2}\left(\frac{M_1M_2}{M}\right)v_{\rm n}^2,
\end{equation}
where $v_{\rm n}$ is given by equation~(\ref{vn}). If $E'_{\rm
  kin}>-E_{\rm grav}$ the new system is unbound. The condition for the
system to be unbound is $v_{\rm n}^2>2v_{\rm orb}^2$. We solve
$v^2_{\rm n}=2v_{\rm orb}^2$ with equation~(\ref{vn}) for the critical
velocity of
\begin{equation}
v_{\rm bound}=-\frac{v_{\rm orb}|\sin \phi|}{\tan i} \pm \sqrt{3 v_{\rm orb}^2+v_{\rm orb}^2 \frac{\sin^2 \phi}{\tan^2 i}}.
\label{vb}
\end{equation}
Because $v_{\rm k}>0$ we take the term with the positive sign. If
$v_{\rm k}>v_{\rm bound}$ then the system is unbound but if $v_{\rm
  k}<v_{\rm bound}$ it remains bound after the supernova kick.

This condition for the system to be bound affects the region in the
$v_{\rm k}$--$\,\phi$ plane that we integrate over to find the
probability distribution for the misalignment angle. To integrate over
all systems, bound or unbound, we integrated equation~(\ref{int}) over
the region $R=R_1+R_2$ shown in Fig.~\ref{contours}. There we also
plot the upper limits on the velocity kicks for the system to remain
bound as dotted lines, $v_{\rm bound}$. Below these lines a system
remains bound but above the kick is too strong and the two stars fly
apart.  The region in which we have bound systems, $R_2$, is much
smaller than the region that can produce the given misalignment angle,
$R_1+R_2$. To find the probability distribution of the misalignment
angle, $P(i)$, for bound systems only we integrate
expression~(\ref{int}) over this smaller region $R=R_2$.

We compute this numerically and plot it in Fig.~\ref{bound} as the
upper dashed line when $v_{\rm orb}=\sigma_{\rm k}$. The lower dashed
line is for $v_{\rm orb}=0.5\sigma_{\rm k}$.  Most kicks unbind the
systems but those that remain bound are somewhat more likely to be
counter-aligned than for smaller kicks.  On the other hand if $v_{\rm
  orb}>\sigma_{\rm k}$ we find that few kicks are able to cause
counter-alignment.

As expected the higher the misalignment angle of a system the lower
the probability of it forming.  We see that, by restricting to only
bound systems, the number with small misalignment is greatly reduced
whereas those closer to counter-alignment are less so.  The
probability of a system closer to counter-alignment than alignment
becomes relatively high for small $v_{\rm orb}$.

\subsection{Mass Loss}

\begin{figure}
  \epsfxsize=8.4cm \epsfbox{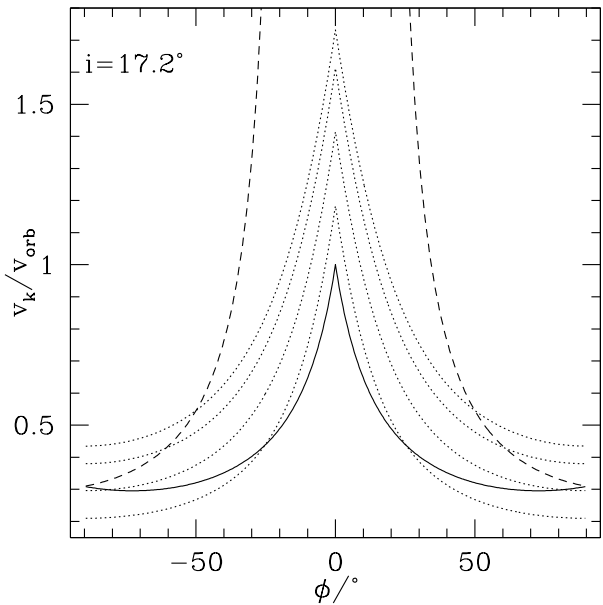}
\caption[]{ This corresponds to the top left panel in
  Fig.~\ref{contours} for $i=17.2^\circ$. The solid line and the
  dashed line are the same in both Figures. The dotted line in
  Fig.~\ref{contours} is the same as the uppermost dotted line here
  and represents the upper boundary for $v_{\rm k}$ which permits the
  post-supernova system to remain bound when no mass is lost. When
  mass is lost from the system the upper limit on $v_{\rm k}$ such
  that the system remains bound decreases as the fraction of mass lost
  $f$ increases. The dotted lines here correspond to $f=$ 0, 0.2, 0.5
  and 0.8.}
\label{contourf}
\end{figure}

So far we have assumed that the mass lost from the system is
negligible. We now allow the mass of star 2 to fall in the supernova
to
\begin{equation}
M_2'=M_2-fM,
\end{equation}
where $f$ is the fraction of mass lost relative to the total mass of
the binary system so that the total mass of the system becomes
\begin{equation}
M'=(1-f)M.
\label{mnew}
\end{equation}
The gravitational energy of the system after the supernova is
\begin{equation}
E'_{\rm grav}=-v_{\rm orb}^{'2}\frac{M_1M_2'}{M'},
\end{equation}
where $v_{\rm orb}'$ is the relative velocity that the stars would have in a
circular orbit of separation $a$ and
\begin{equation}
v_{\rm orb}^{'2}=\frac{G M'}{a}=v_{\rm orb}^2\frac{M'}{M}=(1-f)v_{\rm orb}^2
\label{right}
\end{equation}
because the instantaneous separation $a'=a$. Note that $v'_{\rm orb}$
is the orbital velocity for the equivalent circular orbit while the
orbit itself is eccentric after the supernova.  The new kinetic energy
is
\begin{equation}
E'_{\rm kin}=\frac{1}{2}v_{\rm n}^2\frac{M_1M_2'}{M'}
\end{equation}
and so the condition for a bound system ($E_{\rm kin}'<-E_{\rm grav}'$)
becomes
\begin{equation}
v_{\rm n}^2<2(1-f)v^2_{\rm orb}.
\end{equation}
Thus we find
\begin{equation}
v_{\rm bound}=-\frac{v_{\rm orb}|\sin \phi|}{\tan i} + \sqrt{(3-2f) v_{\rm orb}^2+v_{\rm orb}^2 \frac{\sin^2 \phi}{\tan^2 i}}.
\end{equation}
In Fig.~\ref{contourf} we plot contours of $v_{\rm bound}$ for varying
$f$ with $i=0.3\,{\rm rad}=17^\circ.2$ and $v_{\rm orb}=\sigma_{\rm
  k}$.  The $v_+$ and $v_-$ contours and the top dotted line remain
the same as in the top left plot in Fig.~\ref{contours}.  The more
mass that is lost in the supernova, the lower is the limit on the kick
velocity for a bound system and so the less likely a bound system with
a given inclination becomes.

In Fig.~\ref{bound} for $v_{\rm k}=\sigma_{\rm k}$ we plot the
probability distribution for bound systems for $f=0$ (upper dashed
line), $0.2$ (upper dotted line) and $0.5$ (lower dotted line).  A
larger $f$ increases the likelihood of counteralignment in bound
systems.

\section{Eccentricity Probability Distribution}
\label{sec:ecc}

In the previous section we discussed the effect of particular
supernova kick distributions on the distribution of orbital
misalignments. We now consider what kick distributions are most able
to give rise to the observed Be-star eccentricity distribution.  Given
the sparsity of the data, the large number of free parameters and the
unknown selection effects, we do not attempt to find a best fit to the
periods and eccentricities of Be stars. Rather we look for a kick
distribution consistent with these observations and then examine its
consequences for the distribution of orbital misalignments.

The new semi-major axis of the orbit after the supernova can be found
from
\begin{equation}
v^2_{\rm n}= G M' \left( \frac{2}{a}-\frac{1}{a_{\rm n}}\right),
\label{vn2}
\end{equation}
where $a$, the old semi-major axis, is the instantaneous separation.
Combining this with equation~(\ref{vn}) we can find $a_{\rm
  n}$.  The new system has specific angular momentum
\begin{equation}
\bm{h}'=\bm {r \times v_{\rm n}}.
\end{equation}
where $\bm{r}$ is the separation vector of the stars.  We have
\begin{align}
GM' a_{\rm n} (1-e^2)  = |\bm {r \times v_{\rm n}}|^2 
\end{align}
and so
\begin{align}
GM' a_{\rm n} (1-e^2)  = a^2\left[  v_{\rm k}^2 \sin^2 \phi 
+(v_{\rm k}\cos \omega \cos \phi - v_{\rm orb})^2\right]
\label{ecc}
\end{align}
which can be solved to find the eccentricity, $e$, of the new system
\citep{BP95}.  The binary system is unbound if $e>1$, in which case
$a_{\rm n}\le 0$.  Although we could find eccentricity probabilities
by direct integration in a similar way to the inclinations in the
previous section, it becomes very complicated and we do not learn much
new from the procedure.  Instead we use Monte-Carlo methods to
evaluate the integrals.

We note that $e$ and $i$ both depend only on $v_{\rm k}/v_{\rm orb}$,
$1-f$ and the two angles $\phi$ and $\omega$.  For typical progenitors
of Be~stars the dependence on masses, through $v_{\rm orb}$ and $1-f$
as well as the radius of star~2, turns out to be weak so it can
suffice to concentrate on only one set of masses initially.  We choose
a pre-supernova mass of $M_{\rm 2} = 5\,M_\odot$ that leaves a neutron
star of mass $M_2' = 1.4\,M_\odot$ and a companion mass $M_1 =
15\,M_\odot$.  These masses were used by \citet{BP95} and we have
ensured that we can reproduce their results too.

We use the NAG Library routine {\sc G05CAF} to generate pseudorandom
numbers $\{X_i\}$ uniformly distributed between 0 and~1.  Then for an
isotropic kick distribution
\begin{equation}
\sin \phi = X_{j1}
\end{equation}
and
\begin{equation}
\omega = 2\pi X_{j2}.
\end{equation}
The distribution of orbital periods, $P_{\rm i}$, immediately before the
supernova depends in a complex way on the previous evolution of the
system.  There are many as yet unquantified processes that contribute
to this evolution \citep{H02} and so we stick with the relatively
simple assumption that $\log P_{\rm i}$ is uniformly distributed
between $P_{\rm min}$ and $P_{\rm max}$ so that
\begin{equation}
\log P_{\rm i} = \log P_{\rm i} + X_{j3}(\log P_{\rm max} -\log P_{\rm min}).
\end{equation}
We take $P_{\rm min}$ to be the period at which star~1 would fill its
Roche lobe in a circular orbit if it has the main-sequence radius
given by \citet[ $5\,\rm R_\odot$ for a $15\,\rm M_\odot$
star]{tout1996}. Its Roche-lobe radius $R_{\rm L}$ is approximated by
the formula of \citet{eggleton1983},
\begin{equation}
\frac{R_{\rm L}}{a} = g(q) = \frac{0.49
  q^{\frac{2}{3}}}{0.6q^{\frac{2}{3}} +\log_e (1+q^{\frac{1}{3}})}, \qquad 0< q<\infty,
\end{equation}
where $q=M_1/M_2$.  We take $P_{\rm max} = 10^3\,$d because beyond
this almost all systems are disrupted.

For the post-supernova systems \citet{BP95} rejected any system that
would have filled its Roche lobe if it were circular at its periastron
separation.  So if
\begin{equation}
(1-e)a_{\rm n} <\frac{R_1}{g(q)}
\label{eqperi}
\end{equation}
they rejected the system.  In practice we expect that systems cannot
actually survive down to this separation because tides enforce
pseudosynchronization of star~1 at periastron \citep{hut1991} and so
it ends up spinning up to about 1.16~times faster than it would in a
circular orbit of the periastron separation.  However there is no
equivalent potential theory in the eccentric orbit so we do not try to
be any more precise than condition~(\ref{eqperi}).

We can reproduce figs 4 and 5 of \cite{BP95}. They chose $v_{\rm k}$
to be constant and used a period distribution which is uniform in
$P_{\rm i}$ rather than $\log P_{\rm i}$.  In their figs~4--6, for a
given $x$-axis value, they found the median on the $y$-axis of 10,000
runs and then worked out the regions in which 20, 40, 60, 80
and~$98\%$ of systems lie away from that median. They found very high
values of the mean inclination because they used a high single value
kick velocity.

Instead we integrate over $v_{\rm k}$, distributed according to
equation~(\ref{vk}) up to $850\,\rm km\,s^{-1}$ using Simpson's rule
and $\omega$, $\phi$ and $P_{\rm i}$ by the Monte Carlo method.  In
our figures we prefer to plot contours of probability density in the
2D space normalised so that the probability of lying in the plots is
1.  

Rather than sticking to the fixed masses we distribute the masses of
the companion star from $M_{1\rm min}=5\,\rm M_\odot$ to
$M_{1\rm max}=25\,\rm M_\odot$ according to a mass
function
\begin{equation}
N(M_1) \, dM_1 \propto M_1^{-2.7}\,dM_1
\end{equation}
\citep{kroupa93} which can be generated from
\begin{equation}
M_1=\left( \frac{X_0 - X_{j4}}{k}\right)^{-\frac{1}{1.7}},
\end{equation}
where we find $X_0$ and $k$ from the minimum and maximum masses.
Because the mass range is limited the resulting distributions are not
very different from the fixed initial masses of $M_{\rm 1}= 15\,\rm
M_\odot$ and $M_{\rm 2}=5\,\rm M_\odot$ as used by \cite{BP95} to
represent a typical Be~star binary. Before presenting our results we
discuss the observations with which we compare.

\section{Model Comparison to Observed Systems}

\begin{table*}
\begin{center}
\begin{tabular}{|l|l|l|l|l|l|l|l|}
  \hline
 & Spectral Type& $P_{\rm f}$/d & e &  i\\ \hline
 \span Systems with Emission \\
  \hline 
  0053+604 ($\gamma$ Cas) & B0.5 IVe  & 203.59 & 0.26$^1$ &  about $25^\circ$ $^{43}$  \\
  0115+634 & B0.2 Ve & 24.3 & 0.34$^2$&\\
  0331+530 (BQ Cam) & O8-9 Ve  & 34.3  & 0.3$^3$ & \\
  0352+309 (X-Per) & O9.5 IIIe-B0 Ve &250  & 0.11$^4$ \\
  0535+262 (V725 Tau) & B0 III-Ve & 111 & 0.47$^5$  &  \\
  0834-430 & B0-2 III-Ve  & 105.8  & 0.12$^6$\\
  J1008-57 & O9e-B1e$^{38}$  &247.5  &0.66$^{40}$ \\
  1417-624  & B1 Ve  & 42.12  &0.446$^7$ &\\
  1845-024& Be & 242.18 &0.88$^{37}$ & \\
  J1946+274 & B0-1 IV-Ve & 169.2 & 0.33$^9$ \\
  J1948+32 & B0 Ve & $40.415\pm0.010$ & $0.033\pm0.013$$^{14}$ \\
  2030+375 & B0e & 46.0202 & 0.416$^{10}$\\
  J2103.5+4545 & B0 Ve & 12.66536$\pm$0.00088 & 0.4055$\pm$0.0032$^{11}$ \\
  SAX J0635.2+0533  &     B2V-B1IIIe &    11.2 $\pm$ 0.5& $0.29\pm0.09$$^{12}$  \\
  XTE J0421+560  &   B4 III-V[e] & 19.410 & 0.62$^{13}$     \\     
  4U 2206+543  &      O9.5Ve &    9.570 &  0.15$^{15}$    \\
  B1259-63  & Be  &   1236.72404  & 0.8698869$^{16}$   & greater than $55^\circ$$\,^{42,16}$ \\
 * 0535-668& B2 IIIe$^{35}$ & 16.65 & 0.82$\pm$0.04$^{17}$  \\
  1E 1145.1-6141 &B2Iae &14.365 & 0.20$^{20}$   \\
  GRO J1750-27 &Be &29.817 &0.360$^{36}$ & \\
  LS I +61 303 & B0 Ve& 26.5  & 0.63$\pm$0.11$^{41}$\\
  \hline
  \span Systems without Emission \\ 
  \hline
  1901+03 &OB & 22.58 & 0.036$^8$ &\\
  J0045-7319  &  B  &     51.16926      &      0.80798$^{18}$    & $25-41^\circ\,^{16}$\\
  J1740-3052  &  B $^{19}$ & 231.02965  &  0.5788720 & \\
  2S 0114+650 (LS I +65 010) &    B0.5 Ib&        11.600 &  $0.18\pm 0.05$$^{21}$  \\
 * 4U 1538-52  &   B0 Iab  &       3.730 &   $0.18\pm0.01$$^{22}$  \\     
  4U 1907+09  &     O8-9 Ia &     8.380 &  0.22$\pm$0.05$^{23}$     \\       
  BP Cru (GX 301-2)& B1.5Ia &       41.5 &  0.46$^{24}$  \\  
  OAO 1657-415 & B0-6 Iab & 10.44809$^{25}$ & 0.104$^{26}$  \\
  Vela X-1  & B0.5 Ib   & 8.964368 & 0.0898$^{27}$ \\ 
 * LMC X-4  & O8 III & 1.40841 & 0.006$^{28}$& \\ 
  J1638-4725 &  &      1940.9        &      0.955$^{29}$\\ 
  XTE J1855-026  &O or B&       6.067  & $0.04\pm0.02$$^{32}$ & \\         
  
  \hline
 \end{tabular}
\end{center}

\caption[]{ Binary Be~stars and 
  binary B stars that
  we use in our analysis. We include all the binary B and Be~stars
  we have been able to find reference to in the literature for which both
  periods and eccentricities are reliably known. Most of the data
  that we use comes from the catalogue of Be/X-ray binaries assembled
  by \cite{RP05}. We also include measurements or
  estimates of the misalignment angle $i$ for the few systems for
  which it is available. The systems marked with an asterisk ($\ast$)
  are not included in our analysis because the stellar separation is
  not large enough to accommodate a Be-type disc of size $4\, R_\star$. References for quantities in the Table are given
  $^1$\protect\cite{Harmanec} $^2$\protect\cite{R78}
  $^3$\protect\cite{Stella} $^4$\protect\cite{D01}
  $^5$\protect\cite{Neg00} $^6$\protect\cite{Wilson97}
  $^7$\protect\cite{Finger} $^8$\protect\cite{Galloway05}
  $^9$\protect\cite{Wilson} $^{10}$\protect\cite{Wilson05}
  $^{11}$\protect\cite{Baykal} $^{12}$\protect\cite{K00}
  $^{13}$\protect\cite{B07} $^{14}$\protect\cite{galloway04}
  $^{15}$\protect\cite{Ribo} $^{16}$\protect\cite{Hughes}
  $^{17}$\protect\cite{Hu85} $^{18}$\protect\cite{Kapsi}
  $^{19}$\protect\cite{Stairs} $^{20}$\protect\cite{Ray02}
  $^{21}$\protect\cite{G07} $^{22}$\protect\cite{R06}
  $^{23}$\protect\cite{M84} $^{24}$\protect\cite{VG07}
  $^{25}$\protect\cite{Bild97} $^{26}$\protect\cite{Chak93}
  $^{27}$\protect\cite{Bild97} $^{28}$\protect\cite{levine}
  $^{29}$\protect\cite{Laughlin}
$^{32}$\protect\cite{C02}
  $^{35}$\protect\cite{charles83} $^{36}$\protect\cite{scott97}
  $^{37}$\protect\cite{finger99} $^{38}$\protect\cite{coe94}
  $^{40}$\protect\cite{okazaki01} $^{41}$\protect\cite{cares05}
  $^{42}$\protect\cite{wex98} 
and
$^{43}$\protect\cite{H98}.
}
\label{xray}
\end{table*}

There is now a large number of Be~star binary systems with measured
periods and eccentricities.  The data to which we shall apply our
models are given in Table~\ref{xray}. The bulk of these come from the
catalogue of Be/X-ray binaries assembled by \cite{RP05}\footnote
{http://xray.sai.msu.ru/\~{}raguzova/BeXcat}. We also include O/B stars
that we have found in the literature that have no emission but must
have formed in the same way.  We can use these to look at the
eccentricity distribution and remember that the disc in Be stars may
come and go so that B stars can become Be stars and vice versa.

Because we are primarily interested in Be stars with discs we also
impose the condition that there must be enough room for a decretion
disc of $4\,\rm R_\star$ or so inside the Roche lobe of star~1. We
explain our choice in the appendix but note that there is one Be star
in Table~\ref{xray}, 0535-668, that we do not include in our analysis
because it cannot accommodate such a disc. 

There are three systems for which the misalignment angle $i$ between
the Be~star spin and disc have been estimated. There is one more
system, 28~Tau, which has no reliable orbital parameters but for which
an inclination of $i = 25^\circ - 30^\circ$ has been suggested.

\subsection{Single Maxwellian Peak Velocity Kick Distribution}
\label{single}

\begin{figure*}
  \centerline{\hbox{ \epsfxsize=8cm \epsffile{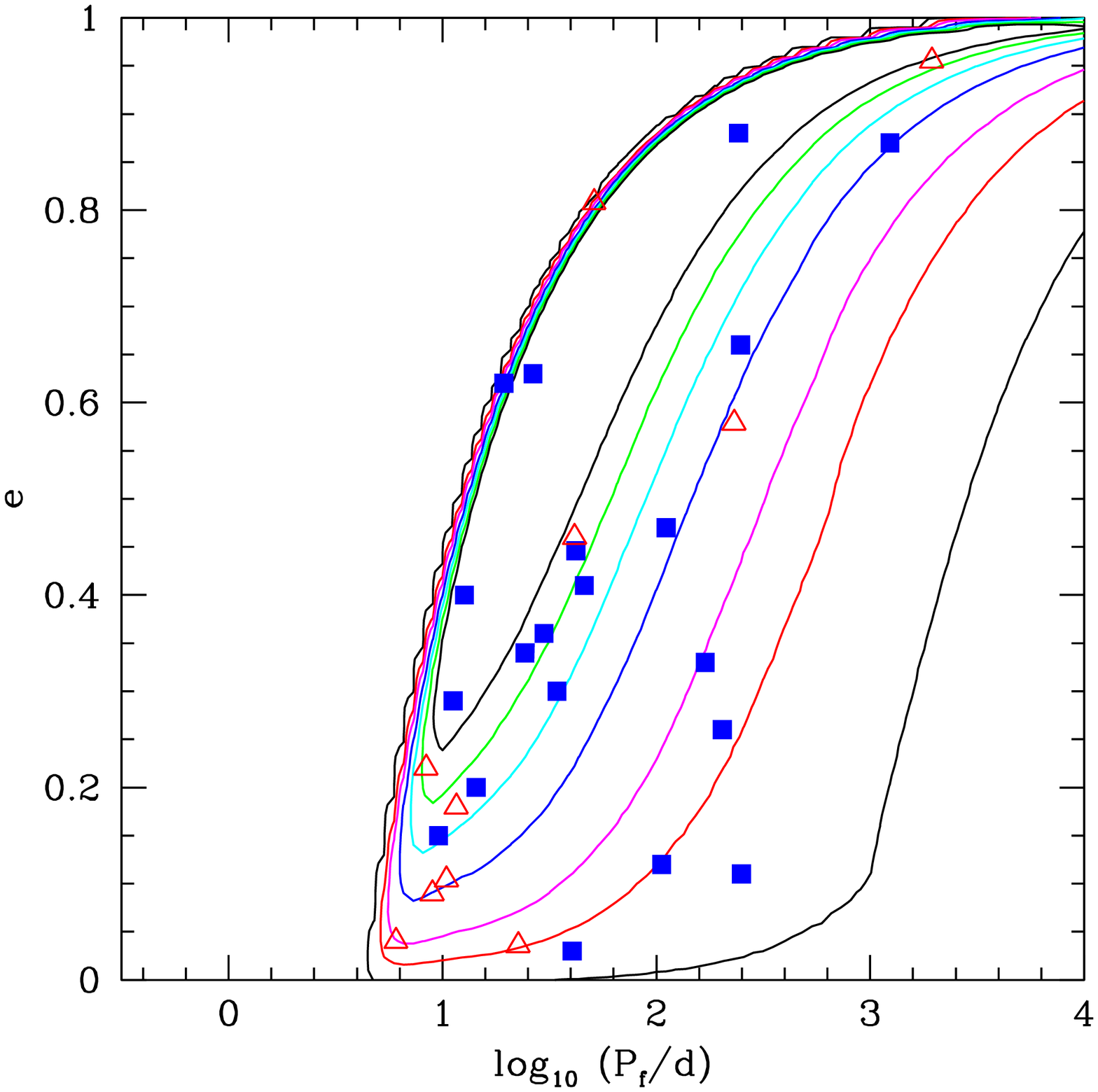}
      \epsfxsize=8cm \epsffile{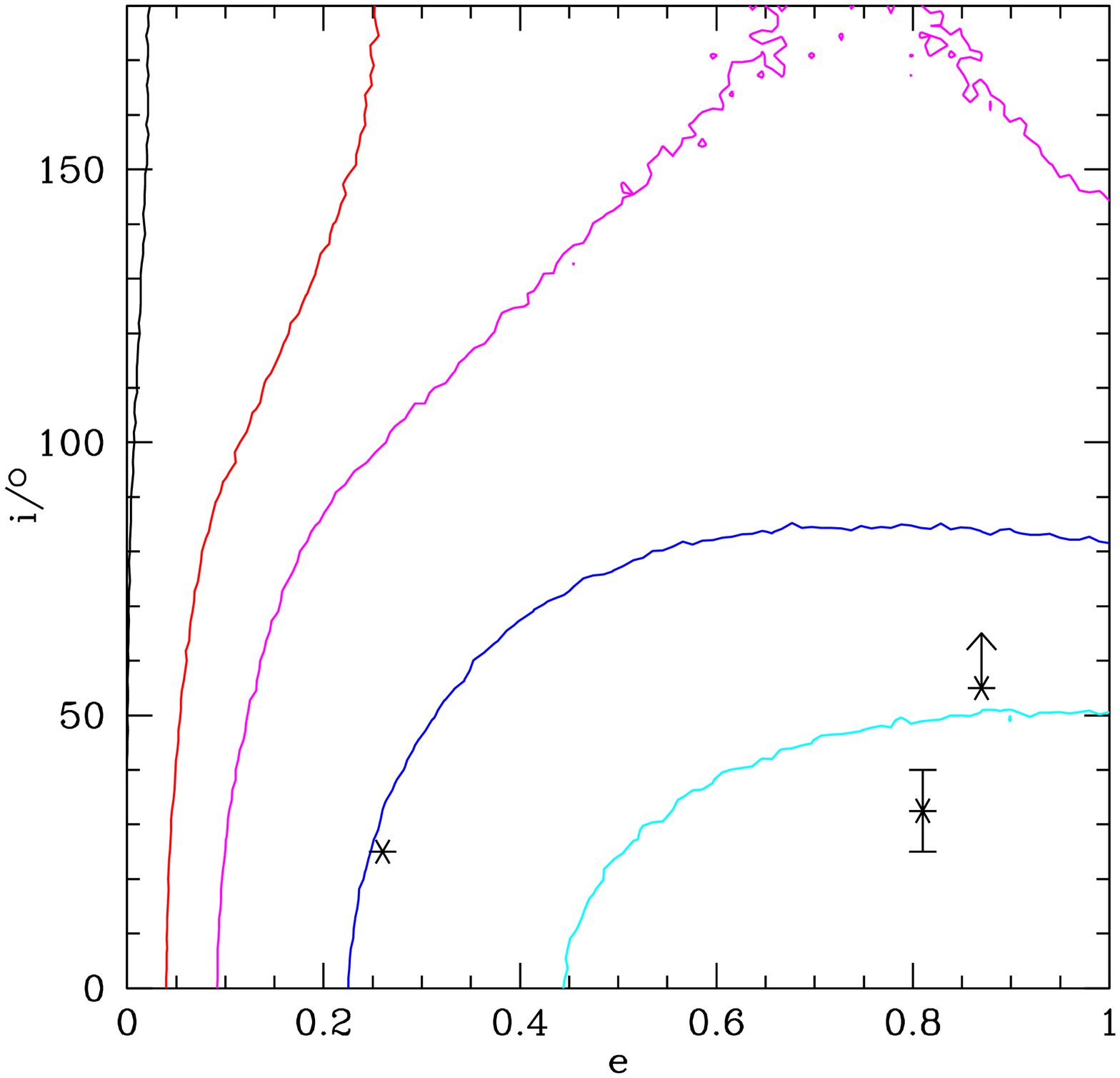} } }
  \caption[]  { Left: the stars listed in Table~\ref{xray} in a
    period--eccentricity diagram. The solid squares are the Be/X-ray
    binaries and the open triangles are the binary B stars.  The
    contours are lines of constant probability density $P(P_{\rm
      f},e)$ when the dispersion of the velocity kick distribution in
    $\sigma_{\rm k}= 265$ km s$^{-1}$ and post-supernova binaries
    which are too tight to permit a disc of size $4 \, R_\star$ are
    excluded (section~\ref{sec:ecc}). The probability density $P$ is
    defined so that the probability of finding a system with $e$ in
    the interval $(e, e+de)$ and with period $\log P_{\rm f}$ in the
    interval $(\log P_{\rm f}, \log P_{\rm f} + d \log P_{\rm f})$ is
    $P \, de \, d \log P_{\rm f}$. The area at small $P_{\rm f}$ is
    excluded by the models.  The outermost contour is at $P = 0.01$.
    Moving inwards the contour levels are $P = $ 0.1, 0.2, 0.4, 0.6,
    0.8 and 1.  Right: contours of equal probability density $P(e,i)$
    for the same models as in the left panel. The contour levels are,
    starting from the left, $P = $ 0.01, 0.1, 0.2, 0.4 and 0.6. The
    asterisks correspond to the systems given in Table~\ref{xray} for
    which there are estimates of the misalignment angle $i$. }
\label{hm}
\end{figure*}

We now consider how well the standard velocity kick distribution of
\cite{hobbs05} fits the observed data. In the left frame of
Fig.~\ref{hm} we plot the probability distribution contours of
eccentricity against final period for these systems and in the right
panel we plot contours in the inclination--eccentricity plane.  The
probability of a highly inclined system is quite small and most
systems end up closer to alignment than to counter-alignment.  Larger
inclinations are more likely for the more eccentric systems.

\begin{figure*}
  \centerline{\hbox{ \epsfxsize=8cm \epsffile{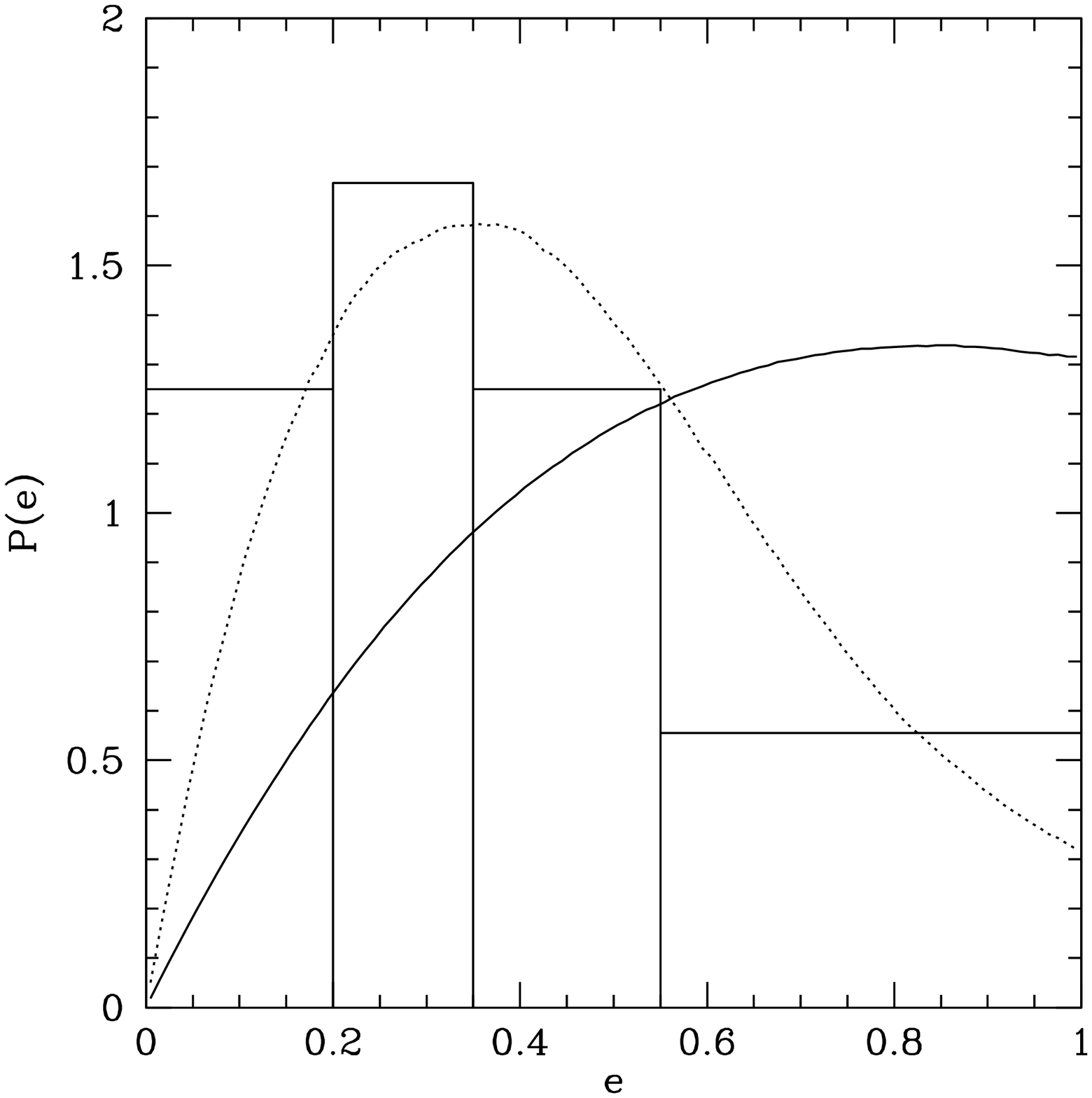}
      \epsfxsize=8cm \epsffile{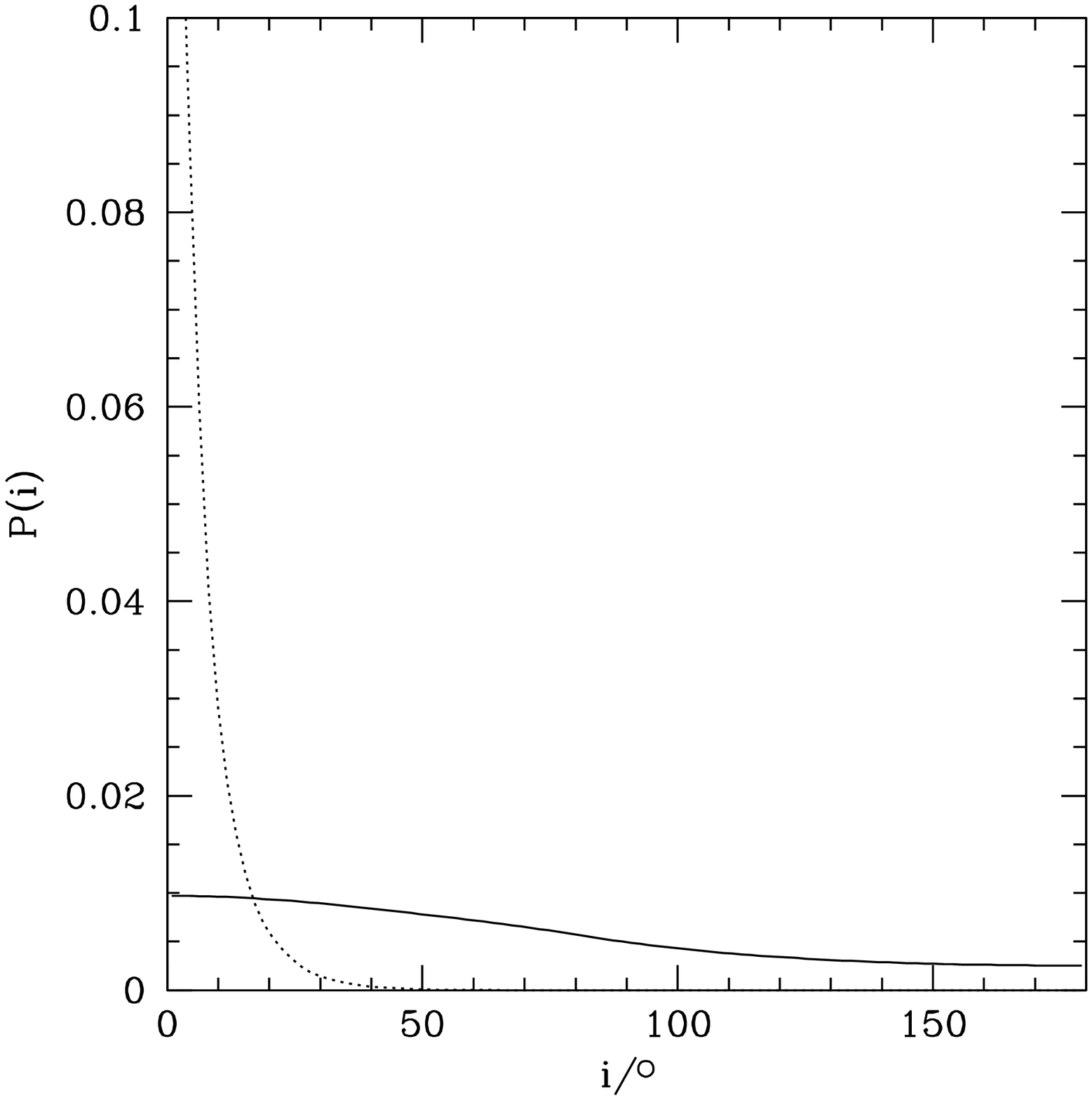} } }
  \caption[] {   Left: the two dimensional probability density $P(P_{\rm f},e)$
    contours, of which are shown in Fig.~\ref{hm} (left), integrated
    over period to give a one dimensional probability density $P(e)$.
    This is plotted as the solid curve (normalised so that $\int_0^1
    P(e) \, de = 1$) and for a kick distribution with $\sigma_{\rm k}
    = 265$ km s$^{-1}$. The dotted line gives the probability density
    $P(e)$ for models constructed with $\sigma_{\rm k} = 15$ km
    s$^{-1}$. The histogram is the observed eccentricity distribution
    for the Be~stars (the solid squares in Figure~\ref{hm}). The lower
    value of $\sigma_{\rm k}$ gives a much better fit to the data.
    Right: as for the left panel but the probability distributions for
    misalignment angles $i$ predicted by the models are plotted. The
    few measured misalignment angles which are known are not plotted
    but all exceed $ i \approx 25^\circ$. It is evident that the solid
    line ($\sigma_{\rm k} = 265$ km s$^{-1}$) is consistent with a
    fairly uniform spread of misalignments while the models with
    $\sigma_{\rm k} = 15$ km s$^{-1}$ (dotted line), which provide a
    better fit to the eccentricity distribution, are not consistent
    with the observed misalignments. For this model only 7 per cent
    have $i > 15^\circ$ and only 1 per cent have $i > 25^\circ$.}
\label{Pi}
\end{figure*}

To illustrate this further, the solid lines in Fig.~\ref{Pi} show the
probabilities of $e$ and $i$ integrated over all systems. We also plot
a histogram of the observed Be~star systems' eccentricities. It is
evident, in line with previous findings on less substantial data sets,
that the \cite{hobbs05} distribution gives a poor fit to the
eccentricity distribution. It does, however, give a fairly flat
distribution of misalignments.  There are too few measured
misalignments to use these as a test but the measured values of are
consistent with the standard \citet{hobbs05} kick distribution.

\begin{figure}
  \epsfxsize=8.4cm \epsfbox{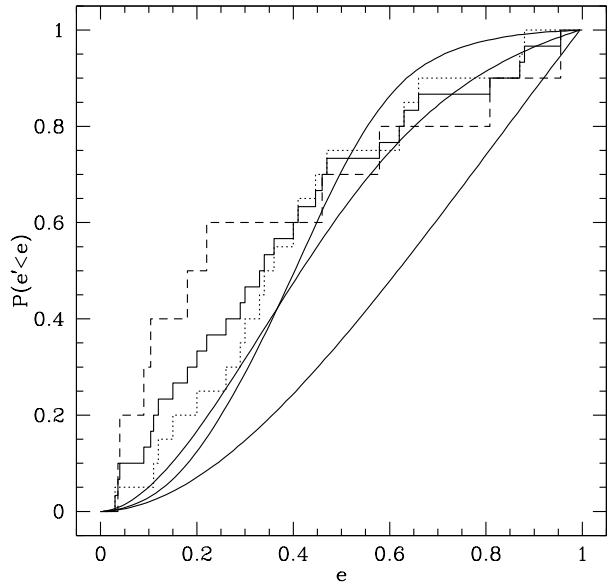}
  \caption[] {Solid lines are the cumulative
    eccentricity probability distribution for $\sigma_{\rm k}=265$,
    $15$, and $5\,\rm km\, s^{-1}$ in order of increasing height on
    the right hand side in the plot. The cumulative eccentricity
    probability distribution for the observed systems is shown for the
    twenty Be stars (dotted line), the ten B stars (dashed line) and
    for the total of thirty B and Be stars (solid line). We use a KS
    test to compare the observed distribution of system eccentricities
    to our model predictions with the results given in Table~\ref{ks}.
    The quality of fit peaks for the Be stars around $\sigma_{\rm
      k}=5\,\rm km\, s^{-1}$ and for the combined Be and B stars at
    around $\sigma_{\rm k}=15\,\rm km\, s^{-1}$.}
\label{cumu}
\end{figure}

Because the final period distribution depends very strongly on our
choice of initial period we have not made use of it for a statistical
comparison with observations except to say that the systems appear to
fit well in Fig.~\ref{hm} (and later in Fig.~\ref{bimodal}). This
implies that our choice of period distribution is reasonable.

We use the Kolmogorov--Smirnov test (KS test) to determine if two
datasets are significantly different. It is non-parametric and
distribution free.  In Fig.~\ref{cumu} we plot the cumulative
eccentricity distributions of the Be stars, of the B stars and of the
B and Be stars combined. We also plot cumulative eccentricity
distribution for our model predictions. We then perform a KS test
between the observed data and our model prediction and give the
results in Table~\ref{ks}. We find the largest deviation of the
observed data from the model and use probability tables for the KS
test to find the probability that the observed sample of stars came
from the distribution predicted by the model.  In Table~\ref{ks} we
give the probability that the observations of Be stars, $P_1$,
B~stars, $P_2$, and combined B~and Be~stars, $P_3$ are consistent with
our various models.  Probabilities smaller than $10^{-4}$ are listed
as zero.  The very small probabilities for the standard \cite{hobbs05}
distribution (the first line in Table~\ref{ks}) demonstrates that it
is essentially impossible that the B and Be~systems have formed as
they are in this way, with a single Maxwellian distribution with
$\sigma_{\rm k}=265\,\rm km\,s^{-1}$.

We cannot, however, immediately rule out a kick distribution in the
form of a single Maxwellian distribution with $\sigma_{\rm k}=265\,\rm
km\,s^{-1}$, because the current eccentricity distribution might not
be representative of the eccentricity distribution immediately after
the supernova.  One possibility is that the systems have begun to
circularise by some mechanism but, given that they are not completely
circular, the timescale on which this circularisation operates is
coincidentally close to the time since the supernova.  We return to
this in Section~\ref{secdisc}.  A second possibility is that there
might be a selection effect on the observed systems.  The interaction
that leads to emission may only occur close to periastron in the very
eccentric systems that spend very little of their time there.
Alternatively it may just be that periods and eccentricities are more
easily measured for the least eccentric systems.  We again return to
this in Section~\ref{secdisc}.

On the other hand the observed eccentricity distribution is consistent
with a Maxwellian kick distribution provided that the kick
distribution is peaked at a lower velocity.  In Table~\ref{ks} we show
the effect of reducing $\sigma_k$ and find very good fits to all
observations when $10 < \sigma_k/{\rm km\,s^{-1}} < 20$. In
Fig.~\ref{Pi} we also plot the dotted lines to show the eccentricity
and inclination distributions for a single Maxwellian peak with
dispersion $\sigma_{\rm k}=15\,\rm km\,s^{-1}$. We see that this curve
appears to fit the eccentricity distribution much better than that
with $\sigma_{\rm k}=265\,\rm km\,s^{-1}$. However, for such low
velocity kicks, the misalignments tend to be small (right panel of
Fig.~\ref{Pi}) and are hard to reconcile with the observed values.

\begin{table}
\begin{center}
\begin{tabular}{|l|l|l|l|l|l|}
  \hline
\multicolumn{4}{|c|}{Single Maxwellian Distribution} \\
  $\sigma_{\rm k}$   & $P_1$ & $P_2$ & $P_3$\\
\hline
265    & 0.0003 & 0.0025 & 0.0000 \\
190     & 0.0003 & 0.0025 & 0.0000  \\
 80     &0.0013 & 0.0035 & 0.0001 \\
 40    & 0.0199 &0.0088 & 0.0057  \\
 20    & 0.1572  &0.0217  & 0.1000  \\
 15      & 0.2550& 0.0238 & 0.1308  \\
 10     & 0.3840& 0.0213  & 0.1141 \\
 5      & 0.4295 & 0.0103 & 0.0445 \\
 2     &  0.3699 & 0.0091 & 0.0378 \\
\hline
\multicolumn{4}{|c|}{Bimodal Maxwellian Distribution} \\
  $\sigma_{\rm k1}$ & $\sigma_{\rm k2}$ & $w_1$  & $P_1$ & $P_2$ & $P_3$\\
\hline
15& 265 & 0.4 & 0.2087 & 0.0217 & 0.1168 \\
15& 190 & 0.5 & 0.0210& 0.0088 & 0.0061 \\
90 & 500 &0.4  & 0.0009& 0.0032 &  0.0000\\
15 & 500 &0.4  & 0.2475 & 0.0229  & 0.1279 \\
\hline
\end{tabular}
\end{center}
\caption[]{ The KS test probability values, the
  likelihood that  the observed systems are chosen from the model distribution.  In
  the upper part of the Table the velocity kick distribution is
  modelled as a single Maxwellian.  In the lower part of the Table the velocity kick
  distribution is modelled as two Maxwellians with 
 $\sigma_{\rm k1}$ and $\sigma_{\rm k2}$ and $w_1$ the relative
  weighting of that with $\sigma_{\rm k1}$ to that with
  $\sigma_{\rm k2}$. The likelihood for the Be stars only is
  $P_1$, B stars is $P_2$ and  for the combined B
  and Be stars is $P_3$.  Probabilities less than $10^{-4}$ appear as
  zero.}
\label{ks}
\end{table}


\subsection{On Axis Kicks}
\label{axis}

\begin{figure*}
  \centerline{\hbox{ \epsfxsize=8cm \epsffile{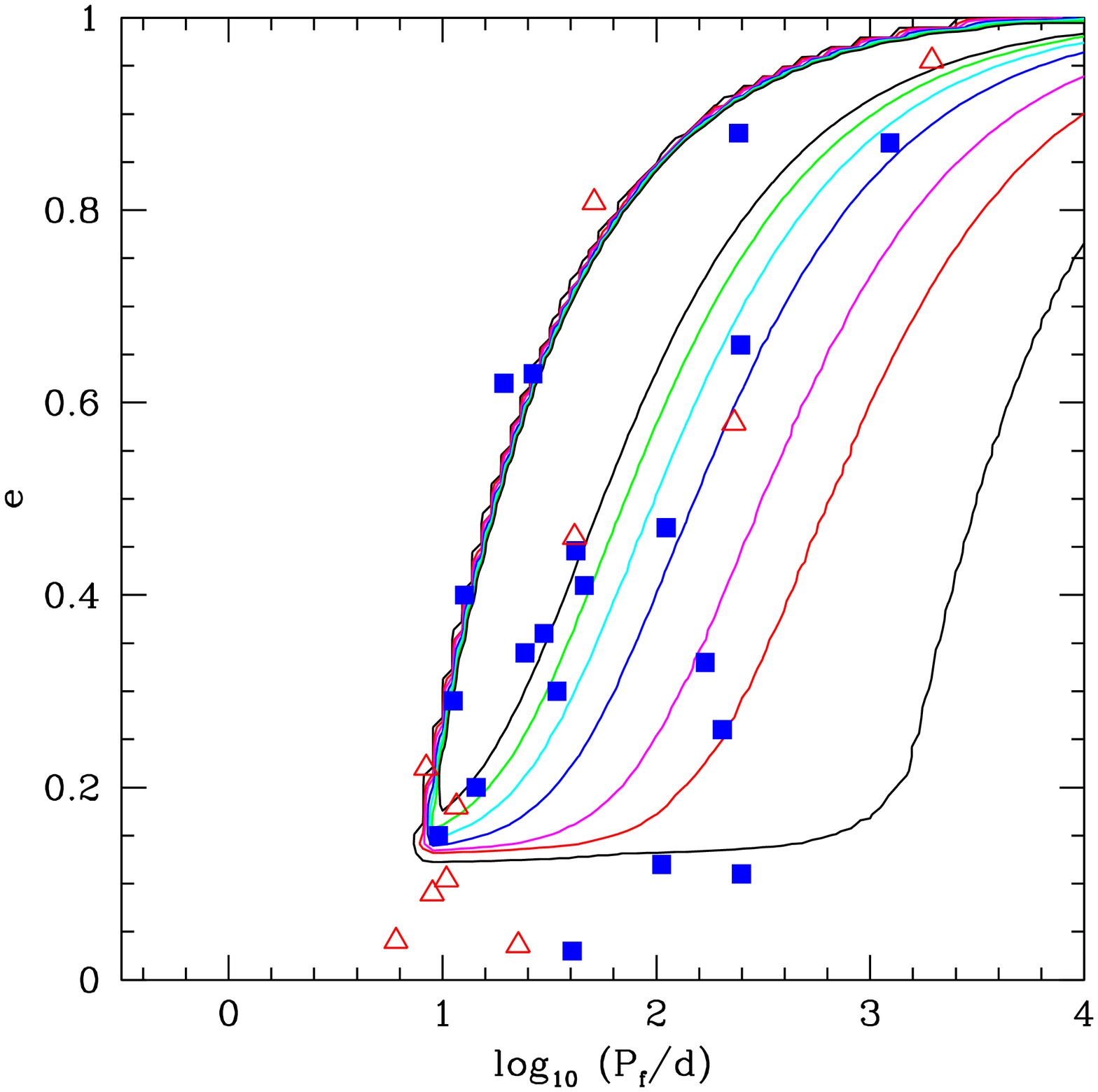}
      \epsfxsize=8cm \epsffile{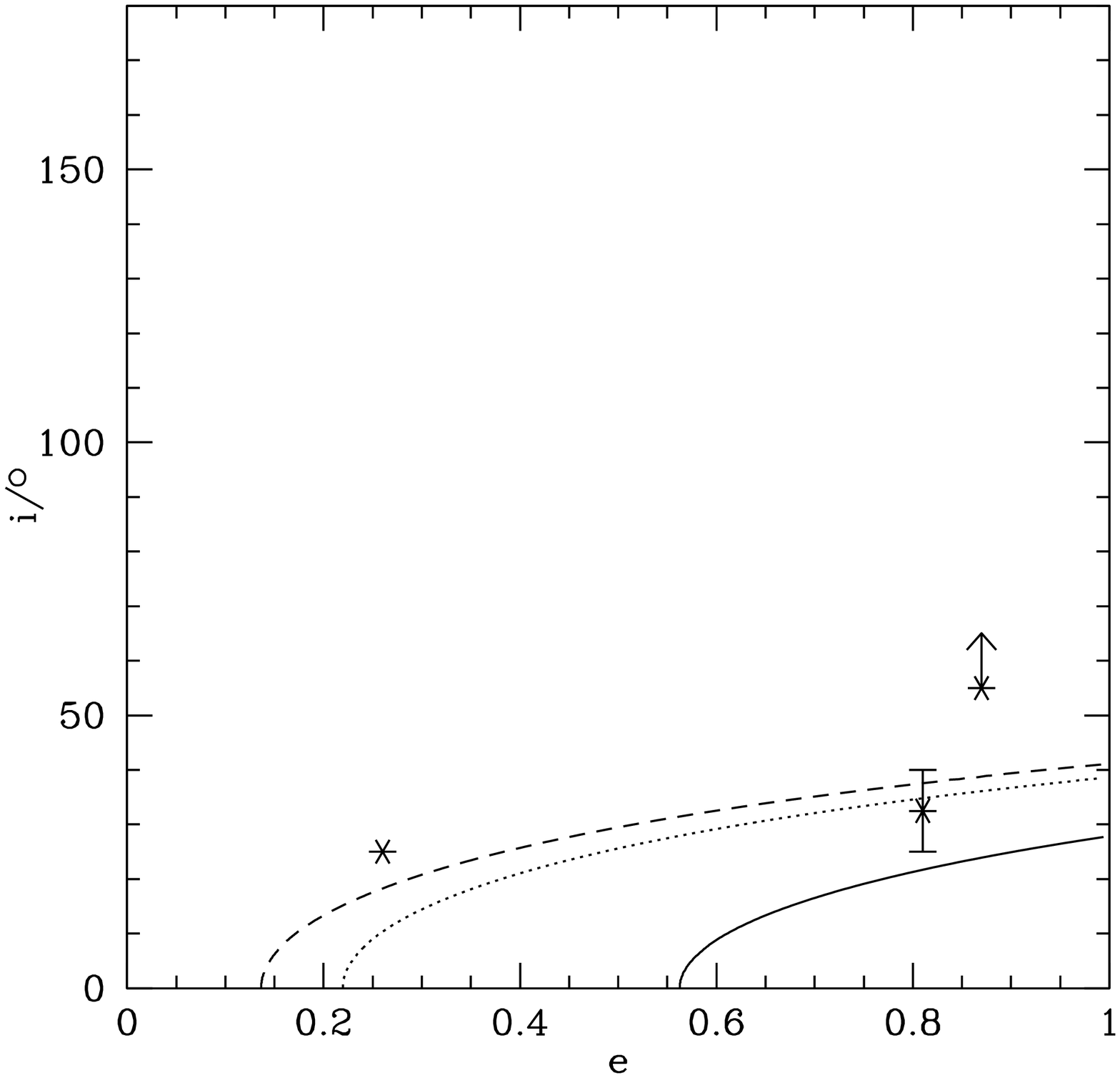} } }
  \caption[] {Left: as for Figure~\ref{hm} except that the direction of the velocity
    kicks are confined to the $z$-axis (the direction of orbital
    angular momentum in the pre-supernova binary). As in
    Figure~\ref{hm} the contours of probability density range from
    $P=0.01$ to $P=1$. The solid squares are the Be~stars and the
    triangles are the B stars without emission from Table~\ref{xray}.
    The models cannot account for the observed low values of the
    eccentricity in some systems.  Right: in this case, because the
    kick direction is fixed, there is a simple relation between
    misalignment angle $i$ and eccentricity $e$ for a given fraction
    $f$ of mass lost in the supernova (equation~\ref{ei}). This
    relation is plotted for $f=0.12$ (solid line), $f=0.18$ (dotted
    line) and $f=0.36$ (dashed line).  The stars show data for the
    systems which have estimated misalignment angles. }
\label{axisfig}
\end{figure*}

It is possible that the direction of the velocity kick in the
supernova is restricted \citep{BP95}. To investigate this we consider
the extreme case that the kick is always directed along the spin axis
of the star, perpendicular to the orbital plane, so that $\phi =
\pi/2$ and $\omega$ is undetermined.  With equation~(\ref{mnew}) and
$v_{\rm orb}^2=GM/a$, we can express equation~(\ref{vn2}) as
\begin{equation}
v_{\rm n}^2=v_{\rm orb}^2(1-f)\left( 2-\frac{a}{a_{\rm n}}\right)
\end{equation}
and then equation~(\ref{ecc}) becomes
\begin{align}
1-e^2=&\frac{1}{1-f} \left( 2-\frac{v_{\rm n}^2}{v_{\rm orb}^2}\frac{1}{1-f}\right) \cr
&\times \left[ \frac{v_{\rm k}^2}{v_{\rm orb}^2} \sin^2 \phi + \left( \frac{v_{\rm k}}{v_{\rm orb}}\cos \omega \cos \phi -1\right)^2 \right].
\end{align}
If the kick lies on the $z$-axis we have $\sin \phi=1$ and so the
misalignment angle between the old and new orbital planes is
\begin{equation}
\cos i = \frac{1}{\sqrt{\frac{v_{\rm k}^2}{v_{\rm orb}^2}+1}}
\end{equation}
and 
\begin{equation}
v_{\rm n}^2=v_{\rm k}^2 +v_{\rm orb}^2.
\end{equation}
We can then relate the eccentricity and the inclination by
\begin{equation}
1-e^2=\frac{1}{1-f}\left( 2- \frac{\sec^2 i}{1-f}\right) \sec^2 i
\label{ei}
\end{equation}
and we plot this in the right panel of Fig.~\ref{axisfig}. For
$M_2=5\,\rm M_\odot$ and $M_2'=1.4\,\rm M_\odot$ we have
\begin{equation}
1-f = \frac{M'}{M} = \frac{1.4 + M_1/{\rm M_\odot}}{5 + M_1/{\rm M_\odot}}.
\end{equation}
For illustration we choose $M_1=5$, $15$ and $25\,\rm M_\odot$ and so $f
= 0.36$,~$0.18$ and~$0.12$.

We plot the eccentricity distributions if the kick is parallel to the
binary orbital axis in the left panel of Fig.~\ref{axisfig}.  We see
that we cannot get highly misaligned systems or low--eccentricity
systems with a kick in the $z$-direction. The low eccentricities could
be explained by circularisation but the highly misaligned systems are
ruled out in this case.

\subsection{Bimodal Distribution of Velocities}
\label{bi}

A bimodal velocity kick distribution has been suggested to explain
both the high-velocity neutron stars and also the fact that neutron
stars appear to be easily contained in globular clusters \citep{katz}.
The escape velocity of a neutron star from a globular cluster is about
$30\,\rm km\, s^{-1}$ and it is generally believed that about 10\% of
neutron stars born within them are retained \citep{druck}.

\cite{arzou} used observed properties of radio pulsars and other
neutron stars to show that a two component velocity distribution fits
the data much better than any one component model. They used a
distribution of velocities with two Maxwellian distributions
\begin{align}
P(v_{\rm k})\, dv_{\rm k}\propto & w_1 \frac{v_{\rm k}^2}{\sigma_{\rm k1}^3}e^{-{v_{\rm k}^2}/{2\sigma_{\rm k1}^2}} \cr
& + (1-w_1)
\frac{v_{\rm k}^2}{\sigma_{\rm k2}^3}e^{-{v_{\rm k}^2}/{2\sigma_{\rm k2}^2}}
 \,dv_{\rm k}
\label{vk2},
\end{align}
where $\sigma_{\rm k1}=90\,\rm km\,s^{-1}$ and $\sigma_{\rm
  k2}=500\,\rm km\,s^{-1}$ and
$w_1=0.4\pm 0.2$ is the weight of the first distribution. We
perform the same KS test on this bimodal distribution and find that
it does not fit our eccentricity distribution well at all (see
Table~\ref{ks}). We note that almost all systems that fall into the
higher peak are disrupted so that the poor fit is entirely due to the
high $\sigma_{\rm k1}$.

We found previously that the best fitting distribution with one peak
had $\sigma_{\rm k}=15\,\rm km\,s^{-1}$. For a bimodal distribution
with $\sigma_{\rm k1}=15\,\rm km\,s^{-1}$ and $\sigma_{\rm
  k2}=265\,\rm km\,s^{-1}$ with equal weight we find that the fit is
somewhat poorer than for the single peak at $\sigma_{\rm k}=15\,\rm
km\,s^{-1}$. However it fits the data significantly better than a
distribution with just $\sigma_{\rm k}=265\,\rm km\,s^{-1}$.  

\begin{figure*}
  \centerline{\hbox{ \epsfxsize=8cm \epsffile{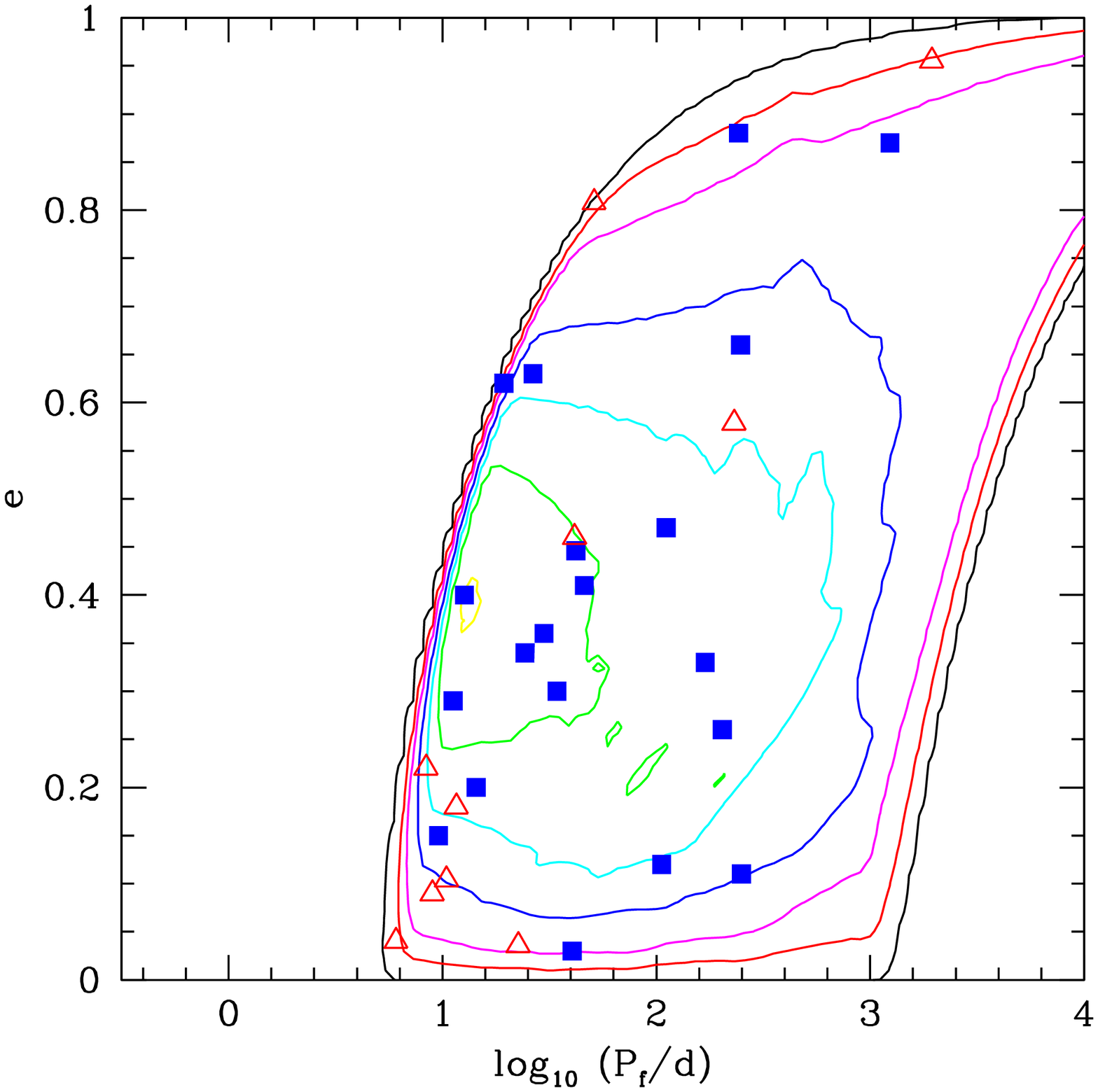}
      \epsfxsize=8cm \epsffile{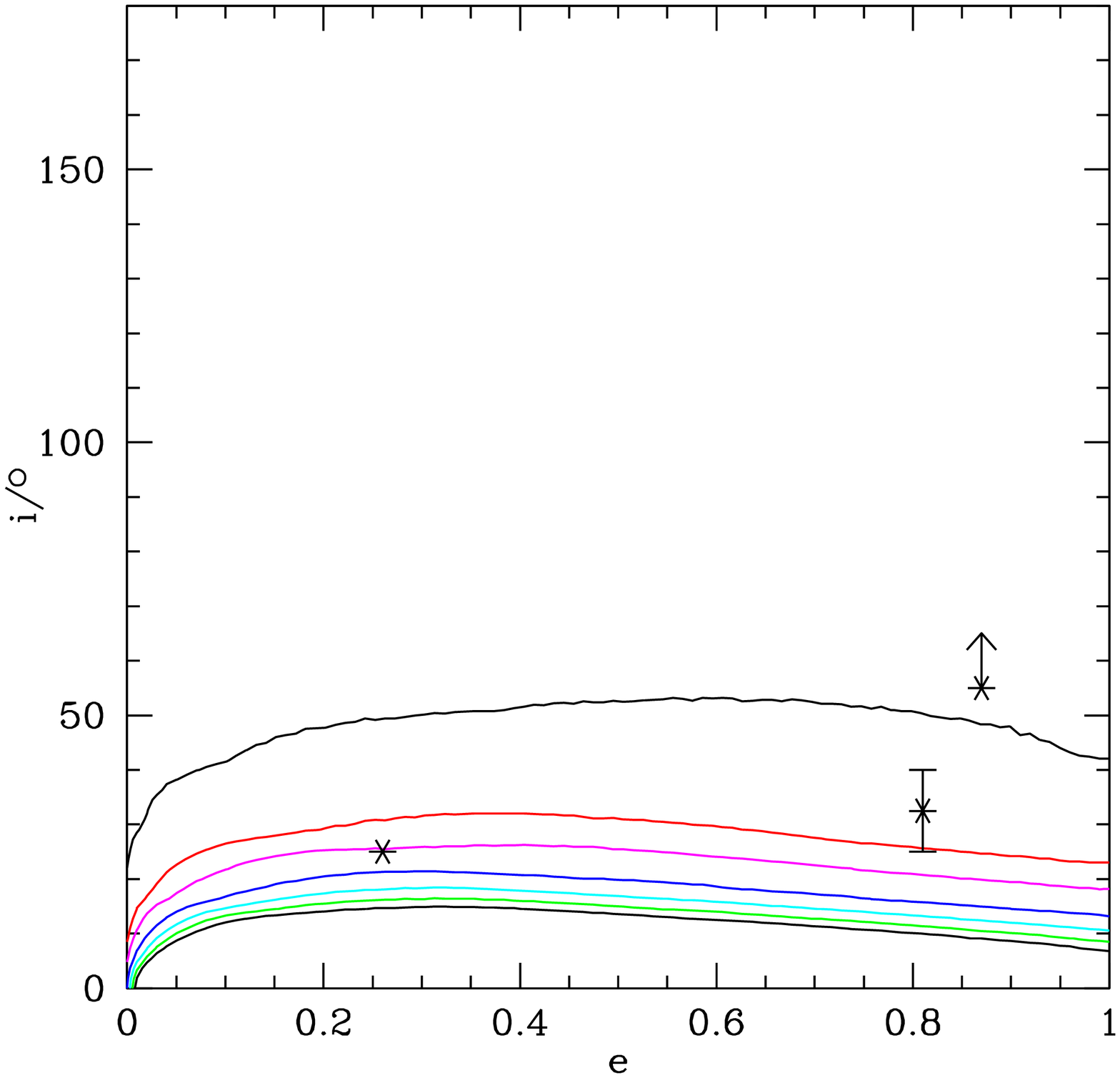} } }
  \caption[] {    As for Fig.~\ref{hm} except that the model velocity kick
    distribution consists of two Maxwellians with $\sigma_{\rm
      k1}=15\,\rm km\,s^{-1}$ and $\sigma_{\rm k2}=500\,\rm
    km\,s^{-1}$ and relative fraction of the former $w_1=0.4$. The
    contours in both plots are as in Fig.~\ref{hm} with contours
    ranging from $P=0.01$ to $P=1$.  Left: the contours of constant
    probability density in the eccentricity--final period plane. Solid
    squares are the Be~stars and triangles are the B stars without
    emission in Table~\ref{xray}.  By comparison with Fig.~\ref{hm} it
    is evident that this model provides a better fit to the
    observations. This is confirmed by the results of the K--S test
    given in Table~\ref{ks}. Right: the probability density in the
    $(e,i)$-plane.  The asterisks are the systems which have an
    observed misalignment angle. The observational points are at low
    values of the probability density, so the observed misalignment
    angles are hard to reconcile with such a model.}
\label{bimodal}
\end{figure*}

The weight factor $w_1$ is unimportant if $\sigma_{\rm k2}$ is large
enough to disrupt most systems.For $\sigma_{\rm k1}=15\,\rm
km\,s^{-1}$ and $\sigma_{\rm k2}=500\,\rm km\,s^{-1}$ with $w_1=0.4$
the fit is as good as the single peak at $\sigma_{\rm k}=15\,\rm
km\,s^{-1}$ (see Table~\ref{ks}).  In Fig.~\ref{bimodal} we plot the
eccentricity--final period and inclination--eccentricity contours for
this velocity kick distribution.  Most of the surviving systems are
from the $\sigma_{\rm k1}=15\,\rm km\,s^{-1}$ part and so the
inclinations are all low.  Nearly all systems with kicks from
$\sigma_{\rm k2}=500\,\rm km\,s^{-1}$ are disrupted.

This bimodal distribution fits the distribution of eccentricities we
find here just because the low-kick systems dominate those that remain
bound.  The effect of the larger kicks is minor simply because most
systems with large kicks are disrupted but this would explain the high
pulsar space velocities. Until we understand the nature of the
supernovae explosions properly in 3-D, bimodal distributions of this
kind will remain good but {\it ad hoc} solutions.

\section{Discussion}
\label{secdisc}

We have modelled the eccentricity distribution of observed B and Be
stars and we confirm the findings of others that the eccentricities
tend to be lower than predicted by a standard \cite{hobbs05} kick
distribution.  The final periods depend on the uncertain choice of
inner boundary and initial period distribution. So, to test the
models, we compared only the probability distribution of the
eccentricities (left panel of Fig.~\ref{Pi}). To quantify this we
performed a K--S test (see Table~\ref{ks}) to see how well the B and
Be stars fit our models. To obtain a good fit from a Maxwellian kick
velocity distribution we require $\sigma_{\rm k}=15\,\rm km\,s^{-1}$,
much lower than the $\sigma_{\rm k}=265\,\rm km\,s^{-1}$ of
\cite{hobbs05}. This can be combined with a second Maxwellian as long
as that has $\sigma_{\rm k}$ large enough to disrupt most systems.  A
combined low and high velocity kicks in the bimodal distribution
can then reproduce the high space velocity pulsars and the
eccentricity distribution of the bound systems.

In Fig.~\ref{Pi} we show that the few estimates of misalignment angles
$i$ between Be~star spin axis and disc axis are too large to be easily
reconciled with kicks with $\sigma_{\rm k}$ as low as $15\,\rm
km\,s^{-1}$. They are much better accommodated by $\sigma_{\rm
  k}=265\,\rm km\,s^{-1}$ (Fig.~\ref{hm}). This is even more true for
the B-star binary PSR J0045-7139 in which the B star spins
retrogradely with respect to the orbit.  However, although the
observed misalignments are indicative of high kick velocities, we
cannot yet base any firm conclusions on the inclinations because (i)
there are only a few, rather uncertain, measurements, (ii) it is
easier to measure a large inclination than a small one and (iii)
misalignment angles of less than the disc opening angle, estimated to
be around 13$^\circ$ \citep{han}, would not be sufficient to easily
give rise to change between Be~star and shell star.

If on the other hand we believe that $\sigma_{\rm k}$ must be larger,
as in the distribution of \citet[][ $\sigma_{\rm k}=265\,\rm
km\,s^{-1}$]{hobbs05} and even the bimodal distribution of \citet[][
$\sigma_{\rm k1}=90\,\rm km\,s^{-1}$, $\sigma_{\rm k2}=500\,\rm
km\,s^{-1}$]{arzou} then we need an alternative explanation of the
lower than expected eccentricities. We first consider whether this
could be due to observational selection effects.  We are less likely
to observe systems with high eccentricity if we only see them when
they are close to periastron. Such systems spend the majority of their
orbital periods closer to apastron and so are less likely to have been
observed. In the case of the Be stars this may simply be due to the
fact that the companion is much more likely to interact with the
decretion disc at periastron. The same would not be true of the plain
B stars which appear to follow a similar eccentricity distribution.
However there are fewer of them and they still may have been selected
by radial velocity variations which would be larger at periastron.

If we assume that we have found all of the Be~star systems in our
lowest eccentricity bin of $0<e<0.105$ (see histogram in
Fig.~\ref{Pi}) then the relative number needs to be brought down by a
factor of about $3.4$.  This would require the actual number of
Be~star systems in our galaxy to be 3.4 times greater than what we
have found and the extra ones must all have high eccentricities. We
have found measured eccentricities and periods for 20 Be systems in
our galaxy and so would need $3.4 \times 20 = 68$ systems to account
for selection effects. There are 67 observed Be~star systems in our
galaxy in the Be/X-ray binary catalogue \citep{RP05} of which 52 do
not have both a measured eccentricity and period. It is unlikely that
the Be stars without observed eccentricities will all turn out to have
high eccentricity. However it is important for observers to measure
the eccentricities of more Be~star systems because this is vital to
rule out selection effects.

Alternatively, and perhaps more interestingly, it is possible that Be
stars did form with an eccentricity distribution favouring large
eccentricities of $e>0.5$ or so. In this case the systems must have
subsequently circularised and the circularisation timescale must be
similar to the lifetime of the Be phase. The periods of Be stars can
be large and the tidal circularisation time, for these stars with
radiative envelopes, is much longer than their lifetimes. One
possibility is that the neutron star interacts with the decretion disc
of the B star at periastron passage.  Such a mechanism might be self
regulating in the sense that the decretion disc only has time to build
up to a large radius when the system is very eccentric and the neutron
star spends a long time far from the B star. So this merits further
investigation particularly if it can be established that supernovae
kicks cannot have such a low dispersion of $\sigma_{\rm k}=15\,\rm
km\,s^{-1}$.

\section{Conclusions}

The distribution of eccentricities in Be and B~stars cannot be
reproduced directly if supernova kicks have a Maxwellian distribution
with $\sigma_{\rm k}>30\,\rm km\,s^{-1}$ or so. Our best fit requires
$\sigma_{\rm k}\approx 15\,\rm km\,s^{-1}$ though this may be combined
to a bimodal distribution with a second Maxwellian with $\sigma_{\rm
  k2}\approx 500\,\rm km\,s^{-1}$, sufficient to disrupt most systems
but to account for the high space velocity pulsars.  We have also
considered the distributions of the misalignment between Be~star spin
and orbit that would result from such a kick distribution. It is
evident that a larger kicks result in larger misalignments. We note
that the data indicate that the low-velocity kicks required to give
the current eccentricity distribution might not be consistent with the
observed misalignments.

If such a low-velocity kick distribution is ruled out then either
selection effects must severely limit the observed distribution to
such an extent that we are only seeing about two in seven of the high
eccentricity systems or the systems must circularise on a timescale
similar to their lifetimes.  Given that this circularisation is not
biased towards low periods, tides in the stars are not sufficient to
be its cause in the wide systems. We postulate that a dissipative
interaction between the neutron star and the decretion disc is a more
likely mechanism. We have suggested that circularisation of Be~stars
might actually be brought about by an interaction between the neutron
star companion and the Be~star's decretion disc. This disc is free to
grow in size while the neutron star is far away. It spends most of its
time at apastron.  Then by periastron the disc may have grown
sufficiently that the neutron star passes through it and is slowed
down. Such an interaction can dissipate energy and thus circularise
the orbit. It cannot however change the angular momentum of the star
nor alter the inclination of the orbit because both these processes
require transfer of angular momentum as well as dissipation of energy.
The moment of inertia of the disc is smaller than that of the B star
and much less than that of the orbit. The star--disc interaction can
therefore easily align the outer parts of the disc with the orbit but
any significant change in the angular momentum of the B star or of the
orbit can only occur on the disc's viscous timescale. Meanwhile
orbital energy can still be dissipated. Thus the misalignment should
still contain information on the kick at formation even if the
eccentricity no longer does.

Inclinations between the pre- and post-supernova orbits are affected
by various factors. Higher kicks lead to fewer bound systems but a
larger probability of counter-alignment. An increase in the mass lost
does the same. Systems with lower pre-supernova periods are less
affected and so tend to be more aligned and more able to survive. For
the standard \cite{hobbs05} kick distribution Be stars ought to be
about three times more likely to end up with aligned rather than
counter aligned orbits but counter-aligned orbits should not be
uncommon. On the other hand if $\sigma_{\rm k}=15\,\rm km\,s^{-1}$ all
orbits should be not far from alignment. In order to distinguish
between these various possibilities more information is required about
the distribution of misalignment angles between the Be~star spin axis
and the orientation of the disc.


\section*{Acknowledgements}
CAT thanks Churchill College for a Fellowship.

\section{Appendix}

Be~stars have a decretion disc around them and theory suggests that Be
stars lose their disc to become B stars. If the disc reappears the
star becomes a Be star again.  If we take this into consideration
there must also be room for the decretion disc within the Roche lobe
of the B~star so we insist that
\begin{equation}
(1-e)a_{\rm n} >C\frac{R_1}{g(q)},
\label{eq57}
\end{equation}
where $C$ is the size of the disc in units of the stellar radius, to
include a system in our analysis.  From interferometry \cite{grund06}
find the size of the disc in $\gamma$ Cas to be $R_{\rm d}=8.1 \pm 1.1
\,\rm R_\star$ where $R_\star$ is the radius of the star so $C=8.1$
and \cite{grund07b} find the disc of X~Per to be about six times
larger than the stellar radius so $C=6$.

\begin{figure}
  \epsfxsize=8.4cm \epsfbox{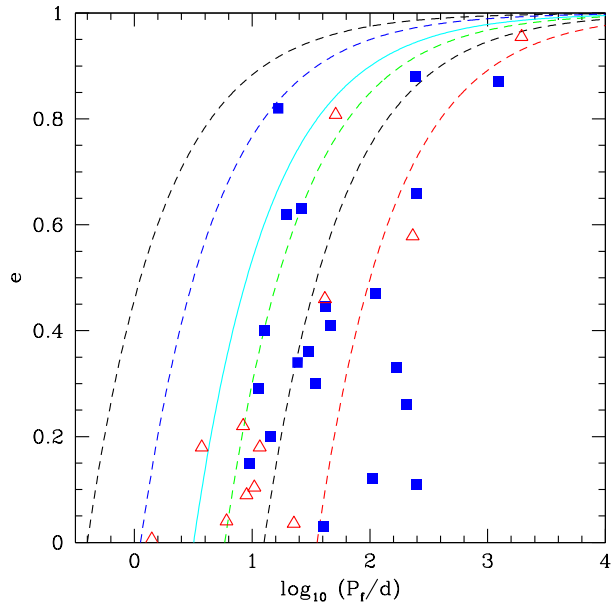}
\caption[]
{ The solid squares are the Be~stars and the triangles are the B stars
  without emission from table~\ref{xray}.  The contours show
  $e=1-a_{\rm min}/a_{\rm n}$ against $P_{\rm f}$ with $C = 1$, $2$,
  $4$ (solid line), $6$, $10$ and $20$ from left to right. The initial
  mass of the exploding star is $M_2=25\,\rm M_\odot$}
\label{peri}
\end{figure}

In Fig.~\ref{peri} we plot the locus defined by equality in
equation~(\ref{eq57}) against $P_{\rm f}$ for $R_1/g(q)=6.9\,\rm
R_\odot$ for $C = 1$, $2$, $4$ (solid line), $6$, $10$ and $20$ in the
case where the initial mass is the maximum of $M_2=25\,\rm M_\odot$.
For each $C$, the area to the left of the contour is invalid.  Only
one Be~star system requires $C$ as low as 1.6.  This system, 0535-668
is in the Large Magellanic Cloud which has low metallicity and hence
stars have smaller radii. The remainder of Be~star systems can be
accommodated if $C=4$. The two B~stars which lie to the left of the
$C=4$ line do not have discs because they show no emission and so we
do not include them in our comparisons.

One effect of increasing $C$ from 0 to 4 is to reduce the number of
high eccentricity systems at low period. Our initial period
distribution biases to low period so this in effect reduces $P(e)$ for
large $e$ in what follows. Thus if we were not to include this our
conclusions in Section~\ref{single} would be even stronger.

\label{lastpage}
\end{document}